\documentclass[pdflatex,sn-mathphys-num]{sn-jnl}
\usepackage[absolute,overlay]{textpos}

\usepackage{soul}

\newcommand{\submissioncitation}{%
\vspace{-21cm}%
\hspace*{-.5cm}%
\begin{minipage}{1.1\textwidth}
\fontsize{10}{10.8}
\fontfamily{ptm}\selectfont
Nurullah Sevim, Mostafa Ibrahim, Sabit Ekin, and Theodore S. Rappaport, "Theoretical Foundations of Waste Factor and Waste Figure with Applications to Fixed Wireless Access and Relay Systems," (Accepted) \textit{npj Wireless Technology}, 2025.
\end{minipage}
\vspace{0.3cm}
}
\usepackage{pdfpages}
\usepackage{graphicx}%
\usepackage{multirow}%
\usepackage{amsmath,amssymb,amsfonts}%
\usepackage{amsthm}%
\usepackage{mathrsfs}%
\usepackage[title]{appendix}%
\usepackage{xcolor}%
\usepackage{textcomp}%
\usepackage{manyfoot}%
\usepackage{booktabs}%
\usepackage{algorithm}%
\usepackage{algorithmicx}%
\usepackage{algpseudocode}%
\usepackage{listings}%

\usepackage{fancyhdr}

\usepackage{mathtools}
\usepackage{array}
\usepackage{hyperref}
\usepackage{cuted}
\usepackage{afterpage}
\usepackage{epstopdf}
\usepackage{subcaption}
\usepackage{float}

\theoremstyle{thmstyleone}%
\theoremstyle{thmstyletwo}%

\theoremstyle{thmstylethree}%

\begin{document}

\title[Article Title]{Theoretical Foundations of Waste Factor and Waste Figure with Applications to Fixed Wireless Access and Relay Systems}

\author*[1]{\fnm{Nurullah} \sur{Sevim}}\email{nurullahsevim@tamu.edu}

\author[2]{\fnm{Mostafa} \sur{Ibrahim}}\email{mostafa.ibrahim@tamu.edu}

\author[1,2]{\fnm{Sabit} \sur{Ekin}}\email{sabitekin@tamu.edu}

\author[3]{\fnm{Theodore~S.} \sur{Rappaport}}\email{tr51@nyu.edu}

\affil*[1]{\orgdiv{Electrical and Computer Engineering}, \orgname{Texas A\&M University}, \orgaddress{\city{College~Station},\state{Texas}, \country{USA}}}

\affil[2]{\orgdiv{Engineering Technology and Industrial Distribution}, \orgname{Texas A\&M University}, \orgaddress{\city{College Station},\state{Texas}, \country{USA}}}

\affil[3]{\orgdiv{NYU Wireless}, \orgname{New York University (NYU)}, \orgaddress{\city{Brooklyn},  \state{New York}, \country{USA}}}

\abstract{

The growing energy demands of next-generation wireless systems call for unified, system-level metrics to evaluate and optimize energy efficiency. This paper advances the concept of the Waste Factor (W), or Waste Figure (WF) in decibel scale, as a general framework for modeling power loss across cascaded communication components. By integrating W into the Consumption Factor (CF)—the ratio of data rate to consumed power—we reveal how component inefficiencies influence the minimum achievable energy per bit. Closed-form expressions are derived for energy-per-bit consumption in both direct and relay-assisted links, along with a decision rule for selecting the more energy-efficient path. While not explicitly modeled, Reflective Intelligent Surfaces (RIS) are shown to fit naturally within this framework. The analysis is further applied to a Fixed Wireless Access (FWA) scenario, where asymmetries in traffic direction and hardware inefficiencies are jointly considered, demonstrating the utility of the Waste Factor in guiding energy-aware system design.
}


\pacs[Accepted at Nature npj Wireless Technology]{July 1st, 2025}



\maketitle

\submissioncitation

\newpage

\section{Introduction}\label{sec1}

{T}{he} world is experiencing a dramatic and accelerating rise in energy consumption. In just the last 40 years, global energy use has doubled—from approximately 90,000 terawatt-hours to 180,000 terawatt-hours annually. Today, humanity consumes around 648 exajoules of energy per year, and this figure is rapidly approaching the zettajoule scale~\cite{iea2025global}. This growth is driven by industrialization, the expansion of digital infrastructure, and the increasing global population. At the same time, energy demand is shifting from traditional sectors such as transportation and manufacturing to data-driven industries like cloud computing, artificial intelligence (AI), and communication networks~\cite{iea2025energyai}. The result is a significant transformation in how, where, and why energy is consumed.

Despite the global push toward renewable energy and improved hardware efficiency, fossil fuels still dominate the energy mix. There is now overwhelming scientific consensus showing a strong correlation between rising carbon dioxide levels and global temperature increases—further reinforcing the need to rethink how energy is consumed and managed across all sectors~\cite{lindsey2024co2}. The ICT (Information and Communications Technology) industry currently consumes around 5\% of global energy, a figure projected to rise to 15–20\% within the next decade~\cite{iea2025global}. AI adoption alone is expected to account for 5–6 exajoules of energy by 2030, driven by massive compute requirements and explosive user demand~\cite{9810097}. With billions of AI queries processed daily—each requiring several watt-hours of energy—the sector’s impact on global energy use is growing exponentially. Moreover, large-scale ICT infrastructure—comprising data centers, access networks, and end-user devices—is becoming one of the fastest-growing energy consumers~\cite{iea_data_centres}.

Wireless communication systems sit at the heart of this transformation. Mobile networks today account for 2–3\% of global electricity consumption, with Radio Access Networks (RANs) responsible for more than 75\% of that energy use \cite{rappaport24waste}. As we transition to 6G (6th Generation) and beyond, wireless systems are becoming significantly more complex and power-hungry. Wide bandwidths, high carrier frequencies, dense cell deployments, massive MIMO configurations, and edge-based AI processing all contribute to rapidly rising energy demands \cite{rappaport2013millimeter,shakya2024propagation,kang2024cellular,shakya2024wideband}.

While hardware innovations, molecular level design choices \cite{11027214} and renewable integration have helped curb the growth of operational energy, reducing energy waste across the network remains a persistent and under-addressed challenge \cite{HuaweiGreen5G}, 
To address the problem of reducing energy waste in a network, several energy efficiency metrics have been proposed, including the power efficiency factor, energy per bit, and the consumption efficiency factor (CEF) \cite{murdock2013consumption,kanhere2022power}. However, these metrics typically provide only a partial view of system performance. 
They often emphasize overall efficiency while overlooking how energy is consumed and wasted across different components in the communication chain. \cite{chen2010energy,xing21high}. Despite efforts by standardization bodies such as European Telecommunications Standards Institute (ETSI), Next Generation Mobile Networks (NGMN), International Telecommunication Union (ITU), and Third Generation Partnership Project (3GPP) \cite{NGMN2021,ETSI2020,ETSI2020_TS103786,ITU2016,ITU2020,3GPP2021}, current frameworks lack a unified, extensible model that can capture energy usage and wastage holistically across diverse wireless architectures.

{Recent efforts, such as the IEEE INGR Energy Efficiency Chapter }\cite{10520551}{, have made significant progress toward modeling energy constraints across the full network ecosystem, identifying key challenges such as base station power, grid limitations, and lifecycle energy impacts. These efforts highlight the importance of a holistic, systems-level view.}

Moreover, \cite{rappaport24waste} proposed the Waste Factor (W) as a unified framework that explicitly partitions RF power into three categories—utilized signal, necessary overhead, and pure waste—across cascaded and parallel stages. By tracking energy from source to sink, W reveals how each amplifier, filter, and beamformer contributes to end‑to‑end inefficiency, enabling actionable optimization in complex wireless chains.  
{Importantly, because W follows power through every RF stage, it inherently captures beamforming gains—e.g., the 20–30 dB array gain achievable at mmWave frequencies—which offsets the $20 \log_{10} (f)$ path‑loss penalty and underpin today’s high-rate 5G Fixed Wireless Access (FWA) links. In free‑space or lightly obstructed channels, those high‑gain arrays actually lose less net power than many sub‑6 GHz systems}\cite{ju2019scattering,6666553,6175397},
{ a fact exploited by carriers rolling out mmWave FWA with massive bandwidths.}

This paper extends W to energy‑per‑bit analysis and applies it to relay‑based networks, such as systems using Reflective Intelligent Surfaces (RIS) setups, as a concrete use case. We then specialize our results to FWA scenarios—where uplink/downlink asymmetry and high‑gain mmWave arrays critically shape overall waste—and, by leveraging previously derived decision rules, we derive guidelines for optimal relay placement and path selection.

The key contributions of this paper are as follows:
\begin{itemize}
    \item \textbf{Extending Waste Factor Analysis}: We build upon the Waste Factor derivations from \cite{rappaport24waste} and apply them to limiting cases in cascaded systems, providing deeper insights into power usage and effects of inefficiencies in multi-stage networks.
    \item \textbf{Recasting Energy-Per-Bit Calculations}: By redefining energy-per-bit metrics through the lens of the Waste Factor, we establish a direct link between energy consumption and waste, offering a more intuitive and comprehensive framework for evaluating efficiency.
    \item \textbf{Optimizing Relay Utilization}: We analyze wireless relay-based systems using Waste Factor principles, identifying conditions under which relays contribute to energy efficiency rather than unnecessary power dissipation.
    \item \textbf{FWA Case Study}: We implement our findings into an FWA scenario considering both uplink and downlink connections and their respective traffic proportions.
\end{itemize}

\section{Results}

\begin{table}[htbp]
\centering
\caption{Summary of Notation}
\begin{tabular}{ll}
\textbf{Symbol} & \textbf{Meaning} \\
\hline
$W$ & Waste factor \\
$WF$ & Waste Figure (in dB scale) \\
$G$ & Gain of a component \\
$G_{\text{ch}}$ & Channel gain \\
$G_{\text{RX}}$ & Gain of the receiver \\
$G_{\text{TX}}$ & Gain of the transmitter \\
$P_S$ & Signal power \\
$P_{NS}$ & Path-related power (excluding signal) \\
$P_{NP}$ & Non-path-related (off-cascade) power \\
$P_{\text{consumed}}$ & Total consumed power \\
$R$ & Data rate \\
$C$ & Capacity of a link \\
$CF$ & Consumption Factor (rate per unit power) \\
$SNR$ & Signal-to-Noise Ratio \\
$N_0$ & Noise power spectral density \\
$E_b$ & Energy per bit \\
$E_{bc}$ & Minimum bit energy consumed by system \\
$W_{\text{TX}}$ & Waste factor of transmitter \\
$W_{\text{RX}}$ & Waste factor of receiver \\
$W_{\text{ch}}$ & Waste factor of the channel \\
$W_{\text{link}}$ & Waste factor of the entire wireless link \\
$d$ & Distance between nodes (e.g., $d_1$, $d_2$, $d_3$) \\
$\alpha$ & Path loss exponent \\
$\rho_u$ & Uplink traffic proportion \\
$\rho_d$ & Downlink traffic proportion \\
$E_3$ & Energy per bit of the direct link \\
$E_{12}$ & Energy per bit of the relay-assisted link \\
$E_\#^{u}$ & Energy per bit during uplink transfer \\
$E_\#^{d}$ & Energy per bit during downlink transfer \\
$P_{NP_\#}^u$ & Non-path-related power during the uplink transfer\\
$P_{NP_\#}^d$ & Non-path-related power during the downlink transfer \\
$C_\#^u$ & Capacity of a link during the uplink transfer\\
$C_\#^d$ & Capacity of a link during the downlink transfer\\
$W_\#^u$ & Waste factor of a link during uplink transfer\\
$W_\#^d$ & Waste factor of a link during downlink transfer\\
$G_{\text{RX,AP}}$ & Receiver gain of access point \\
$G_{\text{RX,BS}}$ & Receiver gain of base station \\
$G_{\text{RX,UE}}$ & Receiver gain of user equipment \\
$W_{\text{TX,UE}}$ & Waste factor of UE transmitter \\
$W_{\text{TX,BS}}$ & Waste factor of BS transmitter \\
$W_{\text{TX,AP}}$ & Waste factor of access point transmitter \\
\end{tabular}
\label{tab:notation}
\end{table}

\subsection{Preliminaries}
Inspired by the concept of the Noise Factor ($F$) introduced by H. Friis in 1944 \cite{friis1944noise}—a foundational metric for quantifying additive noise in cascaded systems—the Waste Factor (W) framework provides a structured approach to evaluating power efficiency in similar architectures. Like the Noise Factor, which captures how noise accumulates across stages of a cascade, the Waste Factor accounts for energy losses at each stage by considering the power levels at the output of every component \cite{ying2023waste}.

To derive the Waste Factor for a cascade, the total power consumed is initially decomposed into three components:

\begin{itemize}
    \item $P_{S}$: The signal power available at the output of an individual stage or at the final output of the cascade.
    \item $P_{NS}$: The power consumed by devices that assist with signal transfer along the cascade path. It includes processing operations, signaling overheads such as pilots and training sequences, coordinating, and others. 
    \item $P_{NP}$: The power consumed by components not on the cascade path and not involved in signal transmission.
\end{itemize}

Based on the previous decomposition, the total consumed power can be expressed as the sum of signal-related and non-path-related power:
\begin{equation}
    P_{\text{consumed}} = \underbrace{(P_{S} + P_{NS})}_{\text{path}} + \underbrace{P_{NP}}_{\text{non-path}}.
    \label{consumed-power}
\end{equation}
Here, $P_{S} + P_{NS}$ represents the power expended directly along the signal path, while $P_{NP}$ accounts for non-signal related power consumption not involved in signal transport.

To quantify the power efficiency of an individual device in the cascade, the Waste Factor is defined as the ratio of path-related power consumption to the output signal power:
\begin{equation}
    W = \frac{P_{\text{path}}}{P_{S}},    
\end{equation}
where $P_{\text{path}} = P_{S} + P_{NS}$ denotes the total power expended along the signal path, including both transmitted signal and supporting internal power. Hence, $W$ is a factor between 1 and $\infty$, and approaching unity is the ideal case.

\begin{figure}
    \centering
\includegraphics[width=0.9\linewidth]{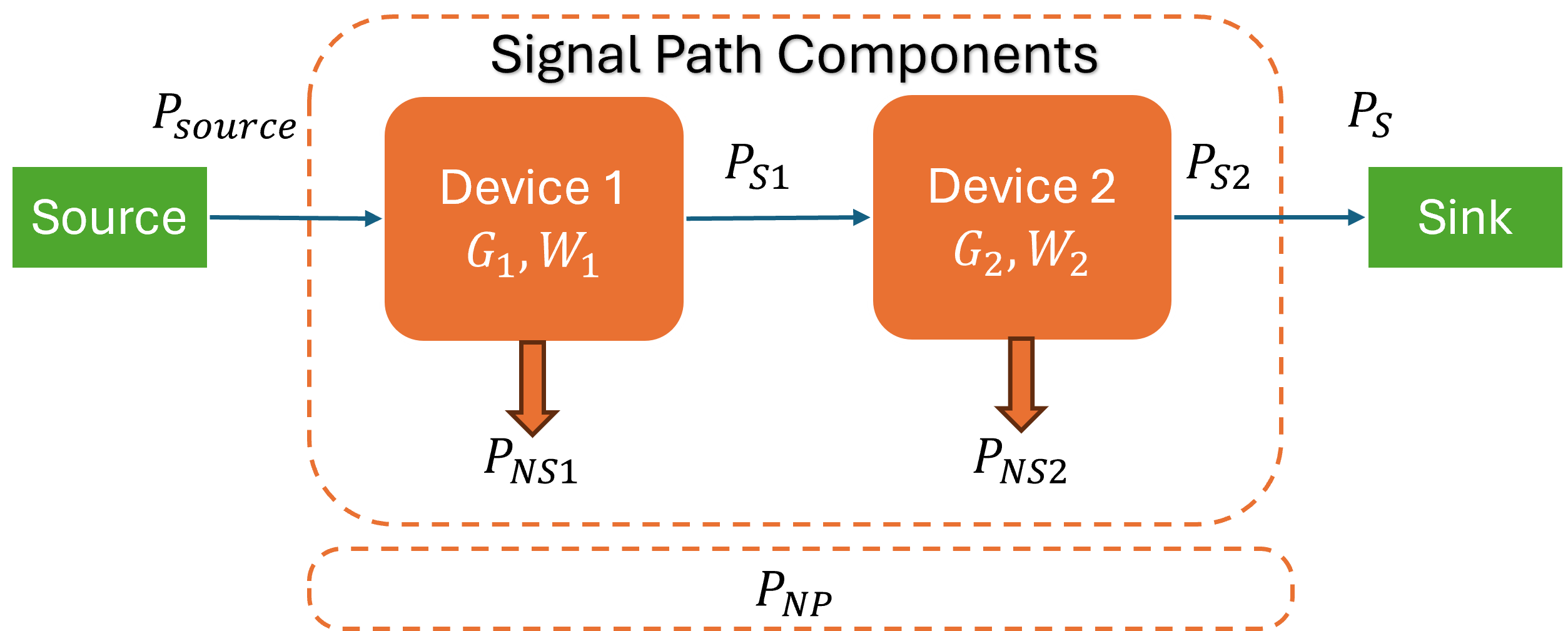}
    \caption{Two device cascade system.}
    \label{cascade}
\end{figure}

To extend the Waste Factor analysis to general cascaded systems, we begin with the simplest case: a cascade composed of two devices, as illustrated in Figure 1. This baseline example lays the foundation for developing a broader formulation applicable to multi-stage cascades.

{To simplify the initial derivation, we focus exclusively on power used along the signal path, i.e., components directly involved in signal transmission and amplification. We temporarily ignore non-path-related power consumption, such as system control processors, idle circuit leakage, thermal management subsystems, and signaling coordination overhead. These elements contribute to the total system energy but do not scale directly with the signal power being transmitted. As such, they are more suitably accounted for in a separate term} (e.g., \( P_{NP} \)) {as done later in Section 2.2 when evaluating the full energy-per-bit cost. While not negligible in practice—especially in high-density or edge-computing scenarios—they are excluded here for analytical clarity in modeling the path-related energy flow.}

\begin{equation}
    P_{\text{path}} = W  P_{S}.
    \label{waste-power}
\end{equation}
Here, $W$ denotes the overall waste factor of the system, and $P_{S}$ is the final output signal power.

The individual power consumption of each device in the cascade can be expressed in terms of their respective waste factors and signal power levels. For the first device, the standalone power consumption is:
\begin{equation}
    P_1 = W_1 P_{S1} - P_{\text{source}}.
\end{equation}
Similarly, for the second device, we have:
\begin{equation}
    P_2 = W_2 P_{S2} - P_{S1}.
\end{equation}
Summing the contributions from both devices and the source, we obtain the total  power spent in the system:
\begin{equation}
    P_{\text{path}} = P_1 + P_2 + P_{\text{source}}.
    \label{total-power}
\end{equation}
Additionally, the output power of the cascade is related to the output of the first device through the gain of the second stage:
\begin{equation}
    P_{S} = P_{S2} = G_2 P_{S1}.
    \label{eqred}
\end{equation}

By substituting the individual power consumption expressions into Eq.~\eqref{total-power}, we obtain:
\begin{equation}
    \begin{split}
        P_{path} &= (W_1 P_{S1} - P_{\text{source}}) + (W_2 P_{S2} - P_{S1}) + P_{\text{source}} \\
          &= W_2 P_{S2} + (W_1 - 1) P_{S1} 
          = \left( W_2 + \frac{W_1 - 1}{G_2} \right) P_S,
    \end{split}
    \label{wf-1}
\end{equation}
where the final step is obtained by using Eq.~\eqref{eqred}. Combining Eqs.~\eqref{waste-power} and \eqref{wf-1} leads to an expression for the overall waste factor of the two-device cascade:

\begin{equation}
    W = W_2 + \frac{W_1 - 1}{G_2}.
    \label{wf-2devices}
\end{equation}

The formulation in~Eq.\eqref{wf-2devices} can be generalized to an $N$-device cascaded system, where the total waste factor is given by:

\begin{equation}
    W = W_N + \frac{W_{N-1} - 1}{G_N} + \dots + \frac{W_1 - 1}{\prod_{i=2}^{N} G_i},
    \label{waste-factor-1}
\end{equation}
or more compactly:
\begin{equation}
    W = \left\{
        1 + \sum_{k=1}^{N} 
        \frac{1}{\prod_{i=k+1}^{N} G_i} 
        \left( W_k - 1 \right)
    \right\}.
    \label{waste-factor-2}
\end{equation}
A similar formulation for evaluating the efficiency of cascaded systems is presented in~\cite{murdock2013consumption}. However, we identified an inaccurate implementation of the formulation from~\cite{murdock2013consumption} in a subsequent study, as detailed in the Supplementary Information section at the end of this paper.

{
To clarify the role of the Waste Factor} \( W \) {relative to other energy-efficiency metrics, we summarize key distinctions below:

- \textbf{Waste Factor}} {\( W \)} {is defined as the ratio of total power expended along a path to the power delivered to the output of a component or cascade stage, i.e.,} \( W = P_{\text{path}}/P_{S} \). {It directly reflects inefficiency due to internal losses and can be recursively applied across cascaded systems. It is agnostic to traffic load or data rate and is suitable for real-time monitoring using power/gain telemetry.}

{- \textbf{Energy per bit} } \cite{shannon1948mathematical} (\( E_b = P/R \)) { captures system-level efficiency, accounting for traffic rate and total power consumed. It reflects real user experience but requires dynamic tracking of traffic and power logs.}

{
- \textbf{Consumption Factor (CF)}} \cite{murdock2013consumption} (\( R/P \)) {is the reciprocal of energy per bit and is also rate-dependent. It is often used in link-level optimization but less modular than }\( W \).

- \textbf{Power Efficiency Factor (PEF)} \cite{razavi2011rf} is typically defined as the ratio of output RF power to total input power in a hardware component, such as a power amplifier: \( \text{PEF} = P_{\text{out}} / P_{\text{in}} \). It is widely used in hardware datasheets to quantify how effectively a device converts DC power into useful RF output. However, PEF is inherently device-specific and does not generalize easily across cascaded systems. 

- \textbf{Power Added Efficiency (PAE)} \cite{sweet1990mic} is another commonly used RF metric, particularly in power amplifier design. It quantifies how efficiently an amplifier converts DC power into additional RF output power, and is defined as:

\begin{equation}
    \text{PAE} = \frac{P_{\text{out}} - P_{\text{in,RF}}}{P_{\text{DC}}}.
\end{equation}

This metric is especially useful for comparing amplifier classes (e.g., Class A, B, AB), but it is limited in scope to amplifier-level evaluations and relies on separating RF input and output power. In prior work \cite{ying2023waste}, this formulation was recast using the waste factor framework, enabling a more general interpretation that integrates naturally into cascaded system models. The use of \( W \) allows PAE-like inefficiencies to be captured in a unified way across a broader range of components and architectures.

- \textbf{Power Usage Effectiveness (PUE)} \cite{dayarathna2015data} originates from data center infrastructure and is defined as the ratio of total facility power to power consumed by IT equipment:

\begin{equation}
    \text{PUE} = \frac{P_{\text{total}}}{P_{\text{IT}}}
\end{equation}

A lower PUE indicates a more energy-efficient data center. While PUE is tailored to large-scale systems, its structure closely resembles that of the waste factor. In fact, PUE is recasted using \( W \) in \cite{ying2023waste}.

{
In contrast to CF and }\( E_b \),{ which require runtime traffic statistics,} \( W \) {enables component-level budgeting and optimization based on static or periodically measured gain and power values. This makes it particularly suited to hardware-aware scheduling and relay placement decisions. Moreover, the derived W, provides a unified and extensible scalar metric that can be applied to any source-to-sink link—whether direct, relayed, passive (e.g., RIS), or hybrid—without needing to reference traffic-specific behavior. This abstraction allows W to serve as a unifying efficiency metric across heterogeneous wireless architectures, enabling consistent energy-aware comparisons and design decisions across layers and technologies.}

To demonstrate the general applicability of the waste factor framework, consider a system composed of two cascaded subsystems, for which the waste factors are already known. Subsystem 1, with waste factor $W_{S1}$ and power gain $G_{S1}$, precedes subsystem 2, which has waste factor $W_{S2}$ and gain $G_{S2}$.

Using Eq.~\eqref{waste-factor-1}, the total waste factor of the entire cascade can be expressed as:
\begin{equation}
    \begin{split}
        W &= \left\{ 1+ \left(W_N -1 \right) + \frac{1}{G_N} \left( W_{N-1} - 1 \right) \right.\\
        &+ \dots + \frac{1}{G_N \dots G_{M+1}} \left( W_{M} - 1 \right)\\
        &\left.+ \dots + \frac{1}{G_N \dots G_2} \left( W_1 - 1 \right) \right\},
    \end{split}
    \label{eq:Wchain}
\end{equation}
where the elements from $1$ to $M-1$ belong to subsystem 1 and elements from $M$ to $N$ belong to subsystem 2 ($1 < M \leq N$).

The waste factors of the two subsystems are:
\begin{align}
    \begin{split}
        W_{S2} = 1 + \left( W_N - 1 \right) 
        + \dots + \frac{1}{G_N \dots G_{M+1}} \left( W_{M} - 1 \right)
    \end{split} \\
    \begin{split}
        W_{S1} = 1 + \left( W_{M-1} - 1 \right) 
        + \dots + \frac{1}{G_M \dots G_{2}} \left( W_1 - 1 \right)
    \end{split}
\end{align}
Then, the overall waste factor can be written as:
\begin{equation}
    \begin{split}
        W &= W_{S2} + \frac{1}{G_{S2}} \left( W_{S1} - 1 \right)= \frac{W_{S2} G_{S2} + W_{S1} - 1}{G_{S2}}.
    \end{split}
    \label{limitcase}
\end{equation}

Looking at Eq.~\eqref{limitcase}, it becomes evident that the waste factor of subsystem 2—typically the final stage in the cascade and closest to the sink—serves as a lower bound for the total system waste factor. This is particularly clear in the limiting case where the first subsystem is ideal, i.e., $W_{S1} = 1$, which yields:

\begin{equation}
    \lim_{W_{S1} \to 1} W = W_{S2}.
\end{equation}

Moreover, Eq.~\eqref{limitcase} reveals a deeper design insight: as the gain of the second subsystem $G_{S2}$ increases, the impact of $W_{S1}$ on the overall waste factor diminishes, and $W$ becomes increasingly dominated by $W_{S2}$. This behavior mirrors classical results from Noise Figure (NF) theory~\cite{noise_figure}, where the noise contribution of the first stage dominates the total NF unless it is followed by a stage with significantly higher gain. 

By analogy, in the context of Waste Factor, this means that power-efficient high-gain stages—especially those closer to the sink—play a critical role in minimizing total energy waste. Just as high-gain low-noise amplifiers (LNAs) are essential for minimizing system noise in receivers, high-gain and energy-efficient power amplifiers are key to achieving overall power efficiency across a cascade. This principle holds for both circuit-level implementations and broader wireless communication architectures.

While transmitters and receivers actively process signals and consume power for amplification, filtering, and computation, the wireless channel plays a fundamentally different role. It serves as a passive attenuator that degrades signal strength over distance due to propagation losses. However, the channel can still be modeled as a cascade node with an associated “waste” behavior, consistent with the system-level treatment of energy flow.

Following the formulation in \cite{rappaport24waste}, the waste factor of such a passive, attenuating component is given by:
\begin{equation}
    W_{\text{ch}} = \frac{1}{G_{\text{ch}}},
    \label{channel-wf}
\end{equation}
where \( G_{\text{ch}} \) is the gain (or attenuation factor) of the wireless channel. 
{Unlike active components such as amplifiers that consume electrical power, the wireless channel does not consume power in a traditional sense. However, it introduces signal attenuation due to propagation losses (e.g., path loss, shadowing, scattering), which leads to a reduction in received signal power. From a system-level perspective, this reduction in signal strength requires the transmitter side to compensate by using more power to provide the desired link quality to the following stage, effectively making the channel appear “wasteful.” Thus, Eq.}~\eqref{channel-wf} {models the channel as a passive component that \textit{incurs} loss rather than consumes power, and the waste factor reflects the additional power that must be expended elsewhere in the system to compensate for this loss. This is explained via example for a passive attenuator using the fundamental definition of Waste Figure, which is then shown to apply to a lossy channel, as well}~\cite{rappaport24waste,ying2023waste}.

Using Eq.~\eqref{waste-factor-2}, the waste factor of the entire wireless link can be expressed as:

{\small
\begin{equation}
\begin{split}
    W_{\text{link}} &= W_{\text{RX}}+ \frac{1}{G_{\text{RX}}} \left( \frac{1}{G_{\text{ch}}} -1 \right) + \frac{1}{G_{\text{RX}} G_{\text{ch}}} \left( W_{\text{TX}} - 1 \right)\\
    &=W_{\text{RX}}+\frac{W_{\text{TX}}}{G_{\text{RX}} G_{\text{ch}}}-\frac{1}{G_{\text{RX}}},
\end{split}\label{waste-factor-link}
\end{equation}
}
where $W_{\text{RX}}$ and $W_{\text{TX}}$ represent the waste factors of the receiver and transmitter, respectively, and $G_{\text{RX}}$ and $G_{\text{ch}}$ denote their corresponding gains.

In scenarios where the product of channel gain and receiver gain is very small, $G_{\text{RX}}G_{\text{ch}}\ll1$ (e.g., due to high path loss and low receiver antenna gain), the waste factor of the entire link becomes very large. In such cases, the expression simplifies to the approximation:

\begin{equation}
    W_{\text{link}}\approx \frac{W_{\text{TX}}}{G_{\text{RX}}G_{\text{ch}}}. 
    \label{approx}
\end{equation}

The approximation in Eq.~\eqref{approx} underscores two key insights. First, to minimize the waste factor in a wireless link, a \textit{high-gain receiver} is critical, as it reduces the burden on the transmitter. Second, while transmitter efficiency is crucial, the waste factor of the receiver also plays an essential role, since—according to Eq.~\eqref{limitcase}—it can form the \textit{lower bound} on the overall link efficiency. However, when the channel gain is very small, the overall waste factor of the link deviates significantly from its lower bound. As a result, the receiver waste factor $W_{\text{RX}}$ does not appear in Eq.~\eqref{approx}.

\subsection{Energy-Per-Bit}

For completeness of analysis, we revisit the energy-per-bit formulation introduced in~\cite{murdock2013consumption} and reformulate the results in terms of the waste factor. Our objective is to incorporate the concept of waste factor into the evaluation of the minimum energy required by a system to transmit a single bit—thereby connecting system-level energy efficiency directly to transmission behavior.

Building on this foundation, we aim to determine the maximum energy-efficient transmission distance that a system can support before energy expenditures become dominated by signal propagation requirements. 

We begin the analysis by leveraging the \textit{Consumption Factor} (CF) introduced in~\cite{murdock2013consumption}, which serves as a general performance metric for evaluating power efficiency in communication systems.
 It is defined as the maximum achievable data rate per unit of consumed power:
\begin{equation}
    CF = \left(\frac{R}{P_{\text{consumed}}}\right)_{max} = \frac{R_{max}}{P_{\text{consumed},min}},
    \label{cf}
\end{equation}
where $R$ is the data rate (in bits per second), and $R_{max}$ denotes the maximum rate the system can support.

To relate CF to fundamental communication limits, consider a general additive white Gaussian noise (AWGN) channel. Based on Shannon’s capacity theorem, the maximum achievable data rate over such a channel is given by:
\begin{equation}
    R_{max} = B \log_2(1 + \text{SNR}),
    \label{capacity}
\end{equation}
where $B$ is the channel bandwidth, and SNR is the signal-to-noise ratio at the receiver. It is important to note that Eq.~\eqref{capacity} holds irrespective of specific modulation or coding schemes.

In AWGN channels, a minimum SNR is required to support a given spectral efficiency $\eta_{sig}=\frac{R}{B}$ (in bps/Hz), and following~\cite{murdock2013consumption} it is defined as:
\begin{equation}
    \frac{\text{SNR}}{M_{\text{SNR}}}=\text{SNR}_{min}=2^{\eta_{sig}}-1,
    \label{msnr}
\end{equation}
where $M_{\text{SNR}}$ is the operating margin above the required minimum $\text{SNR}$.

Now, let us analyze $P_{\text{consumed},min}$. By Eqs.~\eqref{consumed-power} and \eqref{waste-power}, we can write:
\begin{equation}
    \begin{split}
        P_{\text{consumed},min}=P_{NP}+P_{S,min}W\\=P_{NP}+\text{SNR}_{min}P_{\text{noise}}W.
    \end{split}
    \label{pconsumed}
\end{equation}
Eq.~\eqref{pconsumed} introduces the $W$ into the classical Shannon-theoretic framework, explicitly accounting for the additional energy wasted along the path of the signal—not just the power required for ideal transmission. This inclusion, first proposed in~\cite{murdock2013consumption}, provides a more complete picture of real-world power consumption by recognizing inefficiencies present in practical communication systems.

The last equality is obtained by the standard definition of SNR~\cite{sklar2021digital}:
\begin{equation}
    \text{SNR} \coloneqq \frac{P_S}{P_{\text{noise}}}.
\end{equation}

The relationship obtained in Eq.~\eqref{pconsumed} allows us to link CF to the Waste Factor. We can express CF in terms of $\text{SNR}_{min}$ and the system’s waste factor as:
\begin{equation}
    CF = \frac{B \log_2 (1 + \text{SNR})}{P_{NP} + \text{SNR}_{min} P_{\text{noise}}W},
    \label{CF}
\end{equation}
where $P_{NP}$ is the non-path (off-cascade) power consumption. The denominator captures both the waste associated with signal-path power and any wasted energy throughout the network.
The expression in Eq.~\eqref{CF} emphasizes how energy efficiency and waste directly impact the achievable throughput per unit power consumption in wireless systems.

Recall that $\text{SNR}$ can be written in terms of the energy per bit \( E_b \), the data rate $R$, the noise power spectral density \( N_0 \), and the channel bandwidth \( B \) as:
\begin{equation}
    \text{SNR} = \frac{{E_b}{R}}{N_0 B}.
\end{equation}

Now, let us analyze the limiting case where the system bandwidth becomes infinitely large. 
Note that $M_{\text{SNR}}$ in Eq.~\eqref{msnr} becomes $1$ as $B\rightarrow\infty$, i.e., $\lim_{B\rightarrow\infty}\text{SNR}_{min}=\text{SNR}$. From here, we will use $\text{SNR}_{min}$ to denote the signal-to-noise ratio in the limit.
In this regime, recall that we can examine how the data rate $R$ behaves by evaluating the limit of Shannon's capacity formula:
\begin{align}
    \lim_{B\to \infty}R_{max}=\lim_{B\to \infty}B \log_2(1 + \text{SNR}),
    =\lim_{B\to \infty}\frac{\log_2\left(1+\frac{P_S}{N_0B}\right)}{\frac{1}{B}}.
\end{align}

Since both the numerator and denominator approach zero, we apply L'Hôpital's Rule for $\frac{1}{B}$:
\begin{align}
    \lim_{B\to \infty}\frac{\frac{P_S}{N_0}}{1+\frac{P_S}{N_0B}}\frac{1}{\ln{2}}=\frac{P_S}{N_0}\frac{1}{\ln{2}}.
    \label{datarate}
\end{align}

\begin{figure}
    \centering
    \includegraphics[width=\linewidth]{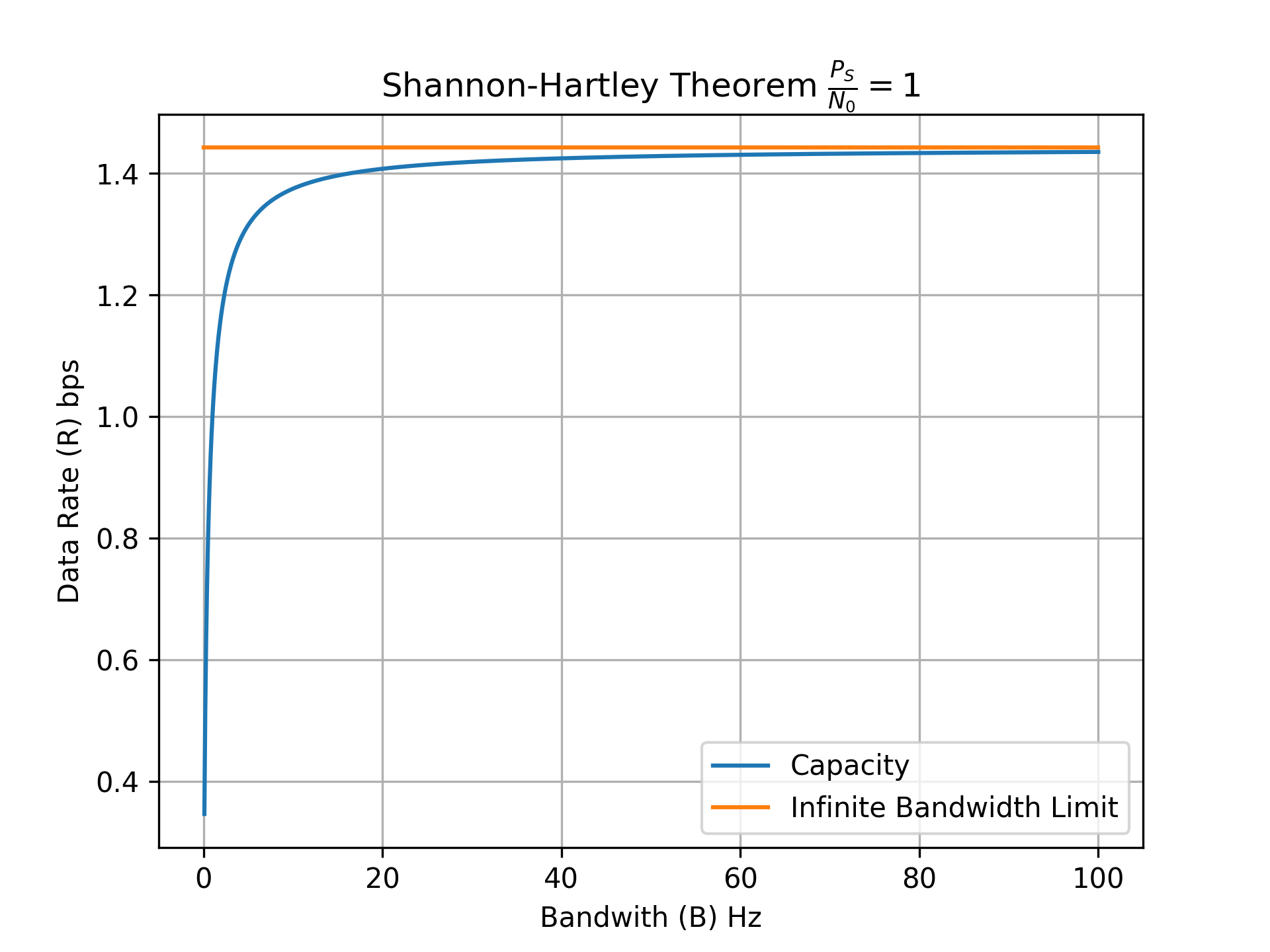}
    \caption{The behavior of the channel capacity $R=B\log_2(1+\text{SNR})$ as bandwidth increases. As $B\rightarrow\infty$, the data rate saturates at its maximum achievable value as given in Eq.~\eqref{datarate}.}
    \label{dataratefig}
\end{figure}

Eq.~\eqref{datarate} shows that the data rate approaches its maximum value of $\frac{P_S}{N_0}\frac{1}{\ln{2}}$ in the wide-band limit. Figure 2 illustrates the behavior of the data rate function from Eq.\eqref{capacity}, confirming that it is a non-decreasing function which asymptotically approaches its upper bound.
Furthermore, since $P_{\text{consumed},min}$ is minimum required power by definition, the minimum energy per bit consumed by the system, $E_{bc}$, becomes the inverse of the CF in the limit of \( B \rightarrow \infty \). We now examine the asymptotic behavior of CF under this observation following~\cite{murdock2013consumption}:
\begin{equation}
    \begin{split}
    \lim_{B \to \infty}CF=\frac{1}{E_{bc}} &= 
    \lim_{B \to \infty} 
    \left\{
        \frac{
            B \log_2 \left( 1 + \text{SNR}_{\text{min}} \right)
        }{
            P_{NP} + \text{SNR}_{\text{min}}  P_{\text{noise}}{W}
        }
    \right\}\\
    & = \lim_{B \to \infty} 
    \left\{
        \frac{
            B \log_2 
            \left( 
                1 + \frac{{E_b}{R}}{N_0 B}
            \right)
        }{
            P_{NP} + 
            \frac{{E_b}{R}}{N_0 B} 
            N_0 B{W}
        }
    \right\}.
    \end{split}
\end{equation}

In the infinite bandwidth regime, the data rate \( R \) approaches the channel capacity limit \( C \). Hence, we simplify the limit as:
\begin{equation}
    \frac{1}{E_{bc}} = \frac{E_b C}{N_0  \ln(2)} \frac{1}{P_{NP} + {E_b}{C}W}.
\end{equation}

Thus the minimum energy per bit consumed by the system becomes:
\begin{equation}
    \begin{split}
        E_{bc} &= \frac{\ln(2) N_0}{E_b}  \frac{P_{NP}}{C} + \ln(2) N_0 W = \frac{P_{NP}}{C} + \ln(2) N_0 W,
    \end{split}
    \label{energy-per-bit}
\end{equation}
where we get the last equality using the Shannon limit $\frac{E_b}{N_0}=\ln(2)\approx -1.59\text{dB}$~\cite{cover1999elements}.

Eq.~\eqref{energy-per-bit} reflects the \textit{true minimum energy per bit} that must be expended by the entire system, relative to the noise spectral density, in order to achieve an arbitrarily low error rate. Importantly, it is distinct from the classical Shannon limit, which considers only the energy within the signal. As a matter of fact, using the Shannon limit we can rewrite Eq.~\eqref{energy-per-bit} as:
\begin{equation}
    E_{bc} = \frac{P_{NP}}{C} + E_b W.
    \label{epb_eb}
\end{equation}

Eq.~\eqref{energy-per-bit} expresses the minimum energy required by a communication system to transmit a single bit, normalized by the noise spectral density, while ensuring an arbitrarily low error rate. The formulation in Eq.~\eqref{energy-per-bit} differs from the classic Shannon limit, which accounts only for the signal energy itself. In contrast, Eq.~\eqref{energy-per-bit} captures the total system-level energy consumption, encompassing both the energy consumed by auxiliary components—represented by $\frac{P_{NP}}{C}$—and the energy wasted across the network during the transmission of a single bit. When the system operates with perfect efficiency—i.e., when no non-path power is consumed (\(P_{NP} = 0\)) and the waste factor is unity (\(W = 1\))—Eq.~\eqref{energy-per-bit} (and therefore Eq.~\eqref{epb_eb}) reduces to Shannon’s limit, implying that all consumed energy is used purely for signal transport, which underscores how the waste factor \( W \) captures the \textit{practical energy overhead} in real-world systems, beyond the idealized channel models.

By substituting the waste factor for a wireless communication link, as derived in Eq.~\eqref{waste-factor-link}, into Eq.~\eqref{energy-per-bit}, we obtain:
\begin{equation}
    \begin{split}
        E_{bc} = 
    \frac{P_{NP}}{C} + \ln(2) N_0 
    \left\{
        W_{\text{RX}} + 
        \frac{1}{G_{\text{RX}}} 
        \left( 
            \frac{1}{G_{\text{ch}}} - 1 
        \right) + \right.
        \left.\frac{1}{G_{\text{RX}} G_{\text{ch}}} 
        \left( W_{\text{TX}} - 1 \right)
    \right\}.
    \end{split}
    \label{1}
\end{equation}

Eq.~\eqref{1} can be further simplified into a more compact form:
\begin{equation}
    \begin{split}
        E_{bc} = 
    \frac{P_{NP}}{C} + 
    \frac{\ln(2) N_0}{G_{\text{RX}} G_{\text{ch}}} 
    \left\{
        G_{\text{RX}} G_{\text{ch}} W_{\text{RX}} + 
        W_{\text{TX}} - G_{\text{ch}}
    \right\}
    \end{split}
    \label{energy-per-bit-waste-factor}
\end{equation}

Recall from Eq.~\eqref{approx} that in scenarios where the channel gain \( G_{\text{ch}}G_{RX} \) is much less than unity, the waste factor of the transmitter becomes dominant. Under the same assumption, the energy-per-bit expression simplifies to:
\begin{equation}
    E_{bc} = \frac{P_{NP}}{C} + \frac{\ln(2) N_0}{G_{\text{RX}} G_{\text{ch}}} W_{\text{TX}}.
    \label{2}
\end{equation}

The expression in Eq.~\eqref{2} reveals two key design insights: First, when the channel gain is low, it is essential to have a high-gain receiver immediately following the channel to prevent excessive power requirements at the transmitter. Second, minimizing the waste factor of the transmitter stage (i.e., the stage before the channel) is critical to ensure energy-efficient transmission. Together, these insights reinforce the importance of system-level design choices in managing energy consumption.

To maximize the total number of bits a communication system can deliver with a fixed energy budget, it is advantageous to minimize the energy cost per bit. One effective strategy to achieve this is to design the system such that non-path power consumption dominates the overall energy per bit. Although this may initially seem counterintuitive, making each bit as cheap as possible allows as many bits as possible to flow through the communication system \cite{murdock2013consumption}.

To formalize this principle, we analyze Eq.~\eqref{energy-per-bit-waste-factor} following~\cite{murdock2013consumption} to determine the maximum allowable link distance at which the non-path power contribution per bit still exceeds the transmission-related power. Beyond this threshold, signal power—amplified to overcome path loss—becomes the dominant contributor to energy per bit.
\begin{equation}
    \frac{P_{NP}}{C} > \frac{\ln(2) N_0}{G_{\text{RX}} G_{\text{ch}}} 
    \left( 
        \left( G_{\text{RX}}{W_{\text{RX}}} - 1 \right) G_{\text{ch}} + W_{\text{TX}}
    \right).
    \label{3}
\end{equation}

Rearranging Eq.~\eqref{3} yields a lower bound on the required channel gain that ensures the non-path power consumption per bit remains greater than the signal-related power expenditure.
\begin{equation}
    G_{\text{ch}} > 
    \frac{\ln(2) N_0 C W_{\text{TX}} }{
        P_{NP} G_{\text{RX}} + N_0 C \ln(2) 
        \left( 1 - G_{\text{RX}}{W_{\text{RX}}} \right)
    }.
\end{equation}

Assuming a standard path loss model where channel gain is given by \( G_{\text{ch}} = \frac{k}{d^\alpha} \), with \( d \) being the link distance, \( \alpha \) the path loss exponent, and \( k \) a constant dependent on system parameters, we can rewrite the inequality as:
\begin{equation}
    \frac{k}{d^\alpha} > 
    \frac{\ln(2) N_0 C W_{\text{TX}} }{
        P_{NP} G_{\text{RX}} + N_0 C \ln(2) 
        \left( 1 - G_{\text{RX}}{W_{\text{RX}}} \right)
    }
\end{equation}

Solving for \( d \), we obtain an expression for the maximum allowable transmission distance that satisfies the energy-efficiency constraint:
\begin{equation}
    \begin{split}
        d < \left\{
        \frac{k}{W_{\text{TX}} \ln(2) N_0 C}
        \left(
            P_{NP} G_{\text{RX}}   
            + N_0 C \ln(2) 
            \left( 1 - G_{\text{RX}}{W_{\text{RX}}} \right)
        \right)
    \right\}^{\frac{1}{\alpha}}.
    \end{split}
    \label{epb-distance}
\end{equation}

The result in Eq.~\eqref{epb-distance} offers important design insights. 
Higher receiver gain \( G_{\text{RX}} \) not only reduces the required transmit power but also increases the feasible communication range.
Also, as the path loss exponent \( \alpha \) increases—indicating a more lossy environment—the maximum sustainable link distance decreases significantly. 

{Furthermore, at mmWave bands the very high array gains achievable at both ends can more than offset the $20\log_{10}(f)$ free‑space loss penalty. Thus, despite larger $\alpha$ in realistic propagation, mmWave FWA systems can realize superior energy‑per‑bit performance—validating the practical advantage of massive‑MIMO beamforming in next‑generation high‑frequency networks.}



\subsection{Waste Factor Analysis in Relay Systems}
\label{wastefactorrelay}
The result in Eq.~\eqref{epb-distance} provides a critical design insight that can be extended to more complex wireless communication systems, particularly those involving relay nodes. Leveraging this foundation, we now investigate {a decision-making mechanism that determines whether to utilize a relay or transmit directly from source to sink, based on energy-per-bit efficiency.} \textbf{This framework is also applicable to modern RIS architectures, which can be viewed as a passive form of relaying. Thus, the insights derived here naturally extend to RIS-assisted links as a special case of relay-based communication.}

\begin{figure}[t]\centering\includegraphics[width=0.5\linewidth]{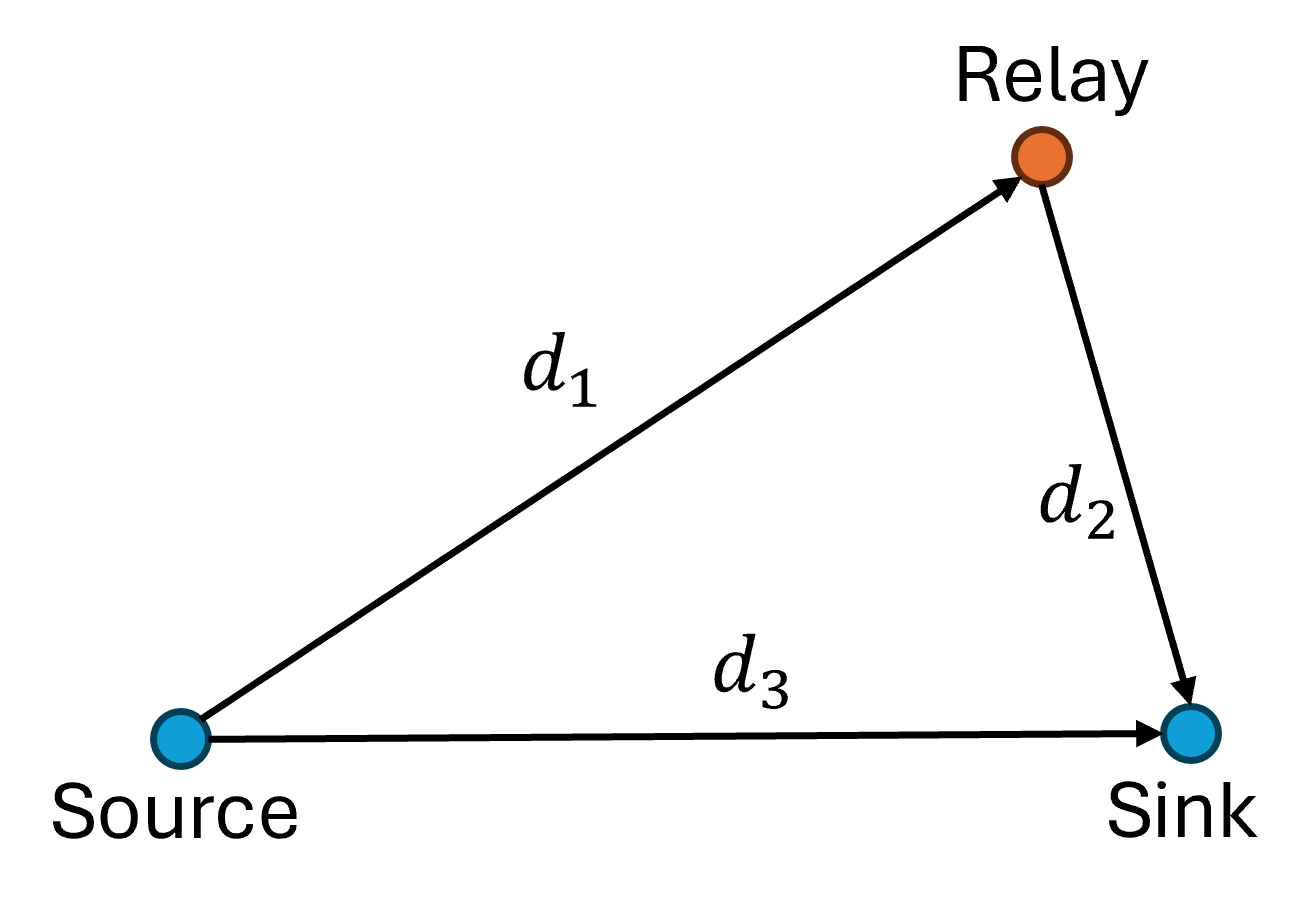}
    \caption{A general scenario where a relay is present in a wireless system.}
    \label{relay-system}
\end{figure}

The general setup is illustrated in Figure 3, where a relay is optionally positioned between a source and a sink.

To evaluate the benefit of using a relay from an energy efficiency perspective, we compare the minimum energy required per bit in two scenarios: direct transmission and relay-assisted transmission.

For the direct link from source to sink, the minimum energy per bit is given by:
\begin{equation}
    E_3 = 
    \frac{P_{NP_3}}{C_3} + 
    N_0 \ln(2)\, W_{3},
    \label{direct-ebc}
\end{equation}
where \( P_{\text{NP}_3} \) is the non-path power consumption and \( C_3 \) is the capacity of the direct channel. From the approximation in Eq.~\eqref{approx}, the waste factor of the direct path can be expressed as:
\begin{equation}
    W_3 \approx \frac{W_{\text{TX,source}}}{G_{\text{RX,sink}} G_3},
\end{equation}
where \( W_{\text{TX,source}} \) is the waste factor of the source transmitter, \( G_{\text{RX,sink}} \) is the gain of the sink receiver, and \( G_3 \) is the direct channel gain.

For the relay-assisted transmission, the energy per bit is computed as the sum of contributions from both hops:
\begin{equation}
    E_{12} = 
    \frac{P_{NP_1}}{C_1} + 
    \frac{P_{NP_2}}{C_2} + 
    N_0 \ln(2) 
    \left(
        W_1 + 
        W_2
    \right),
\end{equation}
where $P_{NP_1}$ and $P_{NP_2}$ are the non-path powers for the two segments, and \( C_1 \), \( C_2 \) their corresponding capacities. The waste factors $W_1$, $W_2$ for source-to-relay and relay-to-sink, respectively, are:
\begin{equation}
    W_1 \approx \frac{W_{\text{TX,source}}}{G_{\text{RX,relay}} G_1}, \label{wd1}
\end{equation}
\begin{equation}
    W_2 \approx \frac{W_{\text{TX,relay}}}{G_{\text{RX,sink}} G_2}, \label{wd2}
\end{equation}
where \( G_1 \) and \( G_2 \) are the gains of the source-to-relay and relay-to-sink channels, respectively.

Now that we have derived expressions for the energy consumption per bit in both relay-assisted and direct transmission scenarios, it is straightforward to establish a decision rule. Assume all non-path power consumptions are equal, i.e., $P_{NP_1} = P_{NP_2} = P_{NP_3} = P_{NP}$, and that all channel capacities are equal to \( C \).

To determine whether the relay path is more energy-efficient, we compare the ratio \( E_{12}/E_3 \). If \( E_{12}/E_3 \) is less than one, then using the relay yields a lower energy cost per bit than the direct link.

Starting with the ratio:
\begin{equation}
    \begin{split}
        \frac{E_{12}}{E_3} = 
    \frac{
        \frac{2 P_{NP}}{C} + N_0 \ln(2) 
        \left( W_1 + W_2 \right)
    }{
        \frac{P_{NP}}{C} + N_0 \ln(2) W_3
    }
    = 
    \frac{
        2 P_{NP} + 
        N_0 C \ln(2) (W_1 + W_2)
    }{
        P_{NP} + 
        N_0 C \ln(2) W_3
    }
    \end{split}
    \label{ratio-eq}
\end{equation}

Imposing the condition \( \frac{E_{12}}{E_3} < 1 \), and solving the resulting inequality, we obtain:
\begin{equation}
    W_3 > \frac{P_{NP} + N_0 C \ln(2)(W_1 + W_2)}{N_0 C \ln(2)}
\end{equation}

Substituting the expression for \( W_3 \) using Eq.~\eqref{approx}:
\begin{equation}
    \frac{W_{\text{TX,source}}}{G_{\text{RX,sink}} G_3} >
    \frac{P_{NP} + N_0 C \ln(2)(W_1 + W_2)}{N_0 C \ln(2)}
\end{equation}

Solving for \( G_3 \), we find:
\begin{equation}
    G_3 < \frac{W_{\text{TX,source}} N_0 C \ln(2)}{G_{\text{RX,sink}} \left( P_{NP} + N_0 C \ln(2)(W_1 + W_2) \right)}
\end{equation}

Assuming a path loss model where the channel gain is given by \( G_3 = \frac{k}{d_3^\alpha} \), and setting \( k = 1 \) for normalization, we obtain:
\begin{equation}
    \frac{1}{d_3^\alpha} < \frac{W_{\text{TX,source}} N_0 C \ln(2)}{G_{\text{RX,sink}} \left( P_{NP} + N_0 C \ln(2)(W_1 + W_2) \right)}
\end{equation}

Rearranging gives:
\begin{equation}
    \begin{split}
        d_3^\alpha > \frac{G_{\text{RX,sink}}}{W_{\text{TX,source}}}
        \left(
            \frac{P_{NP}}{N_0 C \ln(2)} + W_1 + W_2
        \right)
    \end{split}
    \label{distance-1}
\end{equation}

Inserting the expressions for \( W_1 \) and \( W_2 \) from Eqs.~\eqref{wd1} and \eqref{wd2}, and simplifying, we get:
{\small
\begin{equation}
    \begin{split}
        d_3^\alpha &> \frac{G_{\text{RX,sink}}}{W_{\text{TX,source}}}
        \left(
            \frac{P_{NP}}{N_0 C \ln(2)} + 
            \frac{W_{\text{TX,source}}}{G_{\text{RX,relay}} G_1} + 
            \frac{W_{\text{TX,relay}}}{G_{\text{RX,sink}} G_2}
        \right)\\
        &= \frac{G_{\text{RX,sink}}}{W_{\text{TX,source}}}
        \frac{P_{NP}}{N_0 C \ln(2)} + 
        \frac{G_{\text{RX,sink}}}{G_{\text{RX,relay}}} d_1^\alpha + 
        \frac{W_{\text{TX,relay}}}{W_{\text{TX,source}}} d_2^\alpha
    \end{split}
\end{equation}
}

Finally, by assuming no non-path power consumption (i.e., \( P_{NP} = 0 \)) to simplify the analysis and gain insight, we arrive at the clean decision rule:
\begin{equation}
    d_3^\alpha > 
    \left( 
        \frac{G_{\text{RX,sink}}}{G_{\text{RX,relay}}} 
    \right) d_1^\alpha + 
    \left( 
        \frac{W_{\text{TX,relay}}}{W_{\text{TX,source}}} 
    \right) d_2^\alpha
    \label{distance-2}
\end{equation}

Eq.~\eqref{distance-2} provides a geometric condition under which the relay path is more energy-efficient than the direct path, highlighting the trade-offs between relay placement, transmitter and receiver efficiencies, and channel characteristics.

Eq.~\eqref{distance-2} offers valuable insight into how system parameters affect the efficiency of using a relay. The term \( d_1 \), representing the distance over which the relay receives the signal, is scaled by the ratio of the receiver gains at the sink and the relay. Since antenna gain is already incorporated into the normalized path loss model, a relay with a significantly lower receiver gain than the sink could demand a much higher transmission power to compensate, thereby undermining the potential energy savings from a shorter hop.

Similarly, the term \( d_2 \), which denotes the relay’s transmission distance to the sink, is scaled by the ratio of the waste factors of the relay and source transmitters. If the relay’s transmitter is inefficient (i.e., has a high waste factor), then any energy savings achieved by reducing transmission distance may be offset by increased power consumption due to inefficiency.

To build further geometric intuition, consider a free-space propagation model where the path loss exponent \( \alpha = 2 \). In this case, Eq.~\eqref{distance-2} can be recast into the standard form of an ellipse:

\begin{equation}
    1 > \frac{\left(
        \frac{d_1}{d_3}
    \right)^2}{\left(\frac{G_{\text{RX,relay}}}{G_{\text{RX,sink}}}\right)}
     + \frac{\left(
        \frac{d_2}{d_3}
    \right)^2}{\left( \frac{W_{\text{TX,source}}}{W_{\text{TX,relay}}} \right)}
    \label{distance-3}
\end{equation}

Eq.~\eqref{distance-3} defines the interior of an ellipse centered at the origin in normalized \( d_1/d_3 \) and \( d_2/d_3 \) coordinates. As shown in Figure 4, only the shaded region within the ellipse represents valid positions for the relay where it is more energy-efficient to use the relay than to transmit directly. Since all distances must be non-negative, only the first quadrant (where \( d_1, d_2, d_3 \geq 0 \)) is meaningful for deployment decisions.

\begin{figure}
    \centering
    \includegraphics[width=0.9\linewidth]{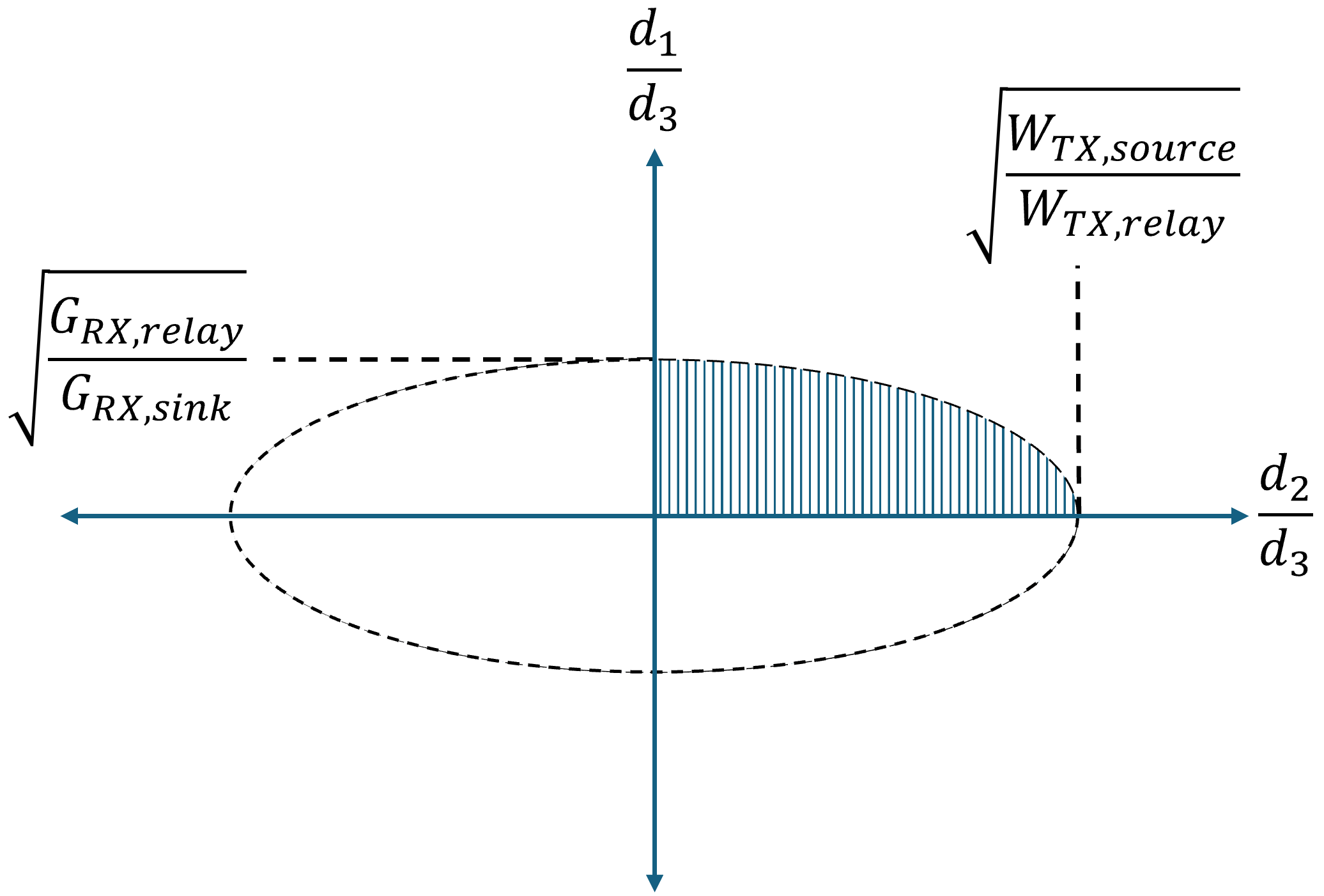}
    \caption{Based on Eq.~\eqref{distance-3}, the striped area denotes the region where it is advantageous to use a relay. In this figure, the path loss exponent is assumed to be 2, the non-path power consumption is assumed to be 0, and $G_{RX}G_{ch}<<1$.}
    \label{ellipse}
\end{figure}

We illustrate several scenarios that highlight the regions where relay placement is advantageous in Figures 5-7. These visualizations are based on the decision rule derived in Eq.~\eqref{distance-2} and offer intuitive insights into how various system parameters influence energy-efficient relay deployment.

In Figure 5, the shaded regions represent different values of the relay transmitter’s waste factor. The configuration assumes a source transmitter with a waste factor of unity, a relay receiver with a gain of 30~dB, a sink receiver with a gain of 10~dB, and a path loss exponent of 5. Each successive region includes all inner regions. As the waste factor of the relay transmitter decreases, the area where using the relay becomes energy-efficient expands significantly.

Figure 6 explores the effect of the path loss exponent. Here, the source transmitter has a waste factor of unity, the relay transmitter has a waste factor of 2, the relay receiver has a gain of 30~dB, and the sink receiver has a gain of 10~dB. Again, each region includes the inner ones. The results show that higher path loss exponents enlarge the zone where relay use becomes beneficial, highlighting the importance of propagation conditions.

Figure 7 investigates the influence of the relay receiver gain. The setup includes a source transmitter with a waste factor of unity, a relay transmitter with a waste factor of 2, a sink receiver with a gain of 10~dB, and a path loss exponent of 6. As the gain of the relay receiver increases, the advantageous relay region also expands, demonstrating the value of high-gain relay components for energy-efficient operation.

Together, these figures validate the analytical insights provided by Eq.~\ref{distance-2}, emphasizing that system-level parameters—such as transmitter waste factors, receiver gains, and environmental propagation characteristics—play a critical role in determining the optimal placement of relays in wireless systems.

{The relay decision rule derived in Eqs. \eqref{ratio-eq}–\eqref{distance-2} assumes equal non-path power consumption ($P_\text{NP}$) and identical channel capacities for the direct and relay-assisted links. The equal non-path power assumption simplifies the analysis and enables closed-form geometric interpretations such as Eq.~\eqref{distance-3}. However, in practical deployments, especially under heterogeneous propagation conditions or when relay nodes employ different processing schemes and form factors, the link capacities and non-path power consumptions may differ significantly. For instance, a relay using complex decoding may have higher $P_\text{NP}$, while its second hop may operate under higher SNR and thus higher capacity. Alternatively, some relay nodes, such as servers or home entertainment systems, may have substantial non-path power as compared to dedicated low-cost or small form factor relays designed specifically and solely for wireless relay functionality. Relaxing the equal non-path power assumption would complicate the derivation but would yield more accurate placement and routing rules. Exploring such generalizations—possibly through numerical optimization or simulation-based models, as well as with AI design approaches, —represents a valuable future direction for this framework.}

\begin{figure}
    \centering
    \includegraphics[width=0.95\linewidth]{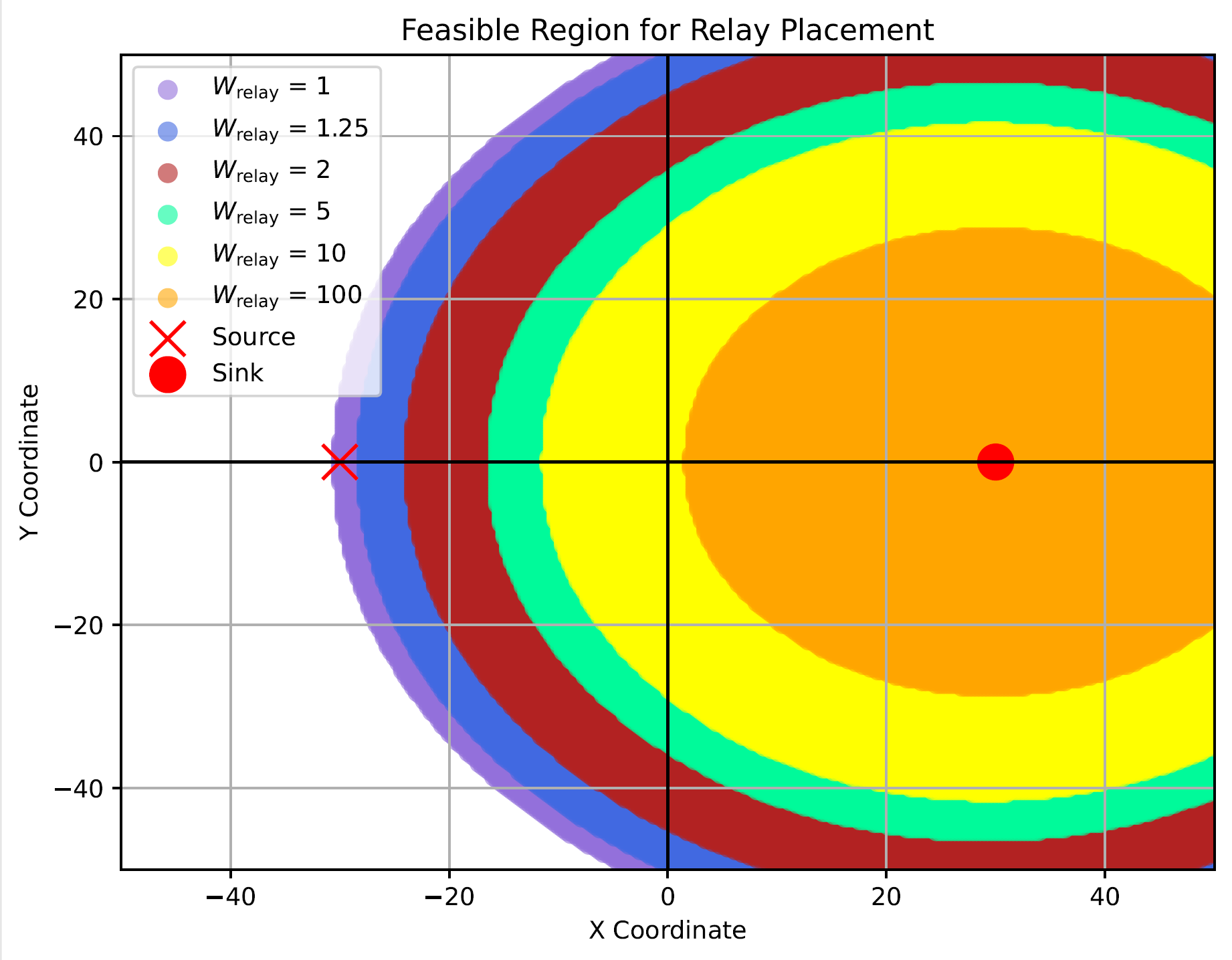}
    \caption{The advantageous regions to place the relay are given in this figure when the waste factor of the relay transmitter is varied. In this setup, the source transmitter has a waste factor of unity, the relay receiver has 30 dB gain, the sink receiver has 10 dB gain, and the path loss exponent is chosen to be 6. Every region includes the inner regions as well.}
    \label{w_vary}
\end{figure}

\begin{figure}
    \centering
    \includegraphics[width=0.95\linewidth]{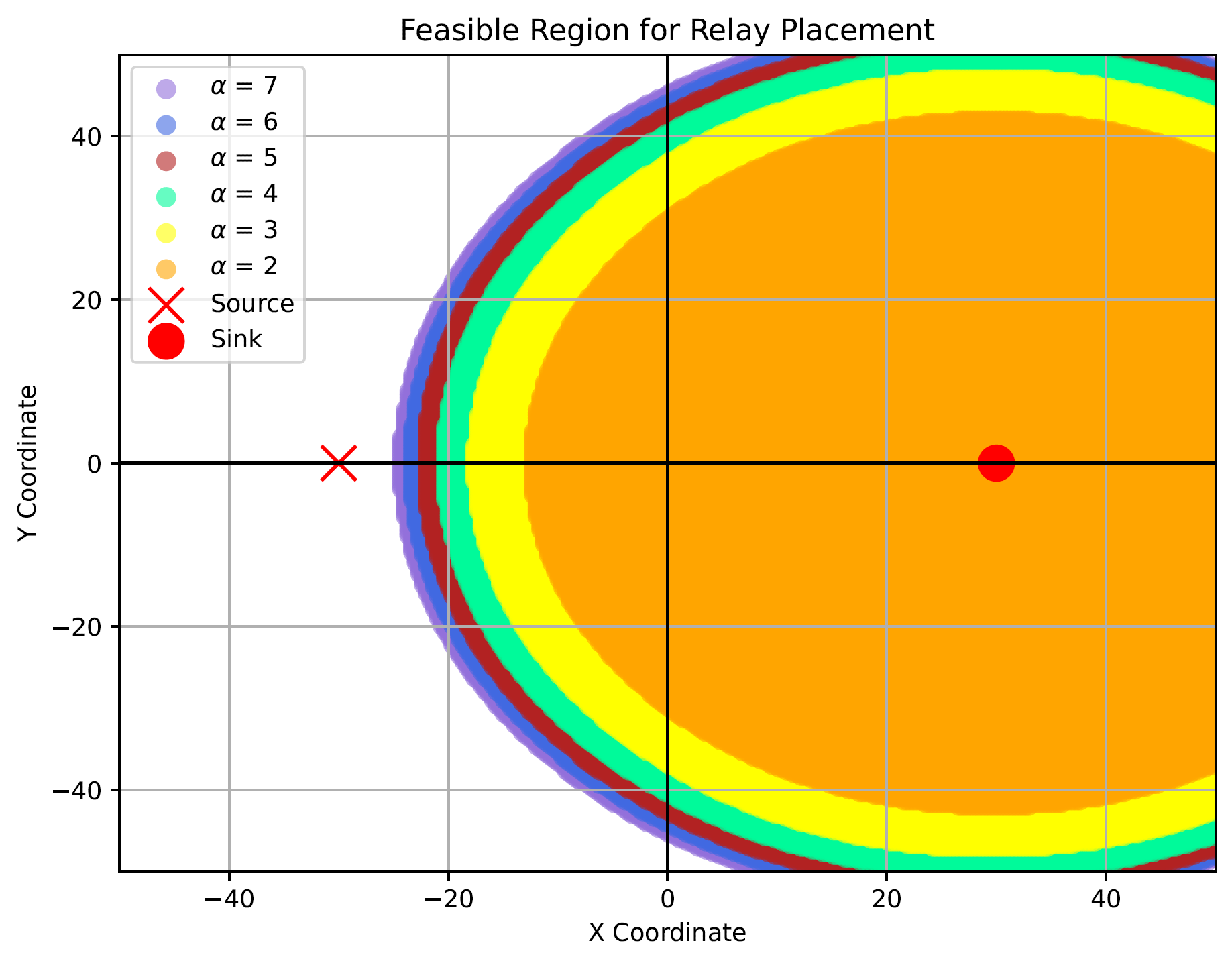}
    \caption{The advantageous regions to place the relay are given in this figure when the path loss exponent is varied. In this setup, the source transmitter has a waste factor of unity, the relay transmitter has a waste factor of 2, the relay receiver has a 30 dB gain, and the sink receiver has a 10 dB gain. Every region includes the inner regions as well.}
    \label{alpha_vary}
\end{figure}

\begin{figure}
    \centering
    \includegraphics[width=0.95\linewidth]{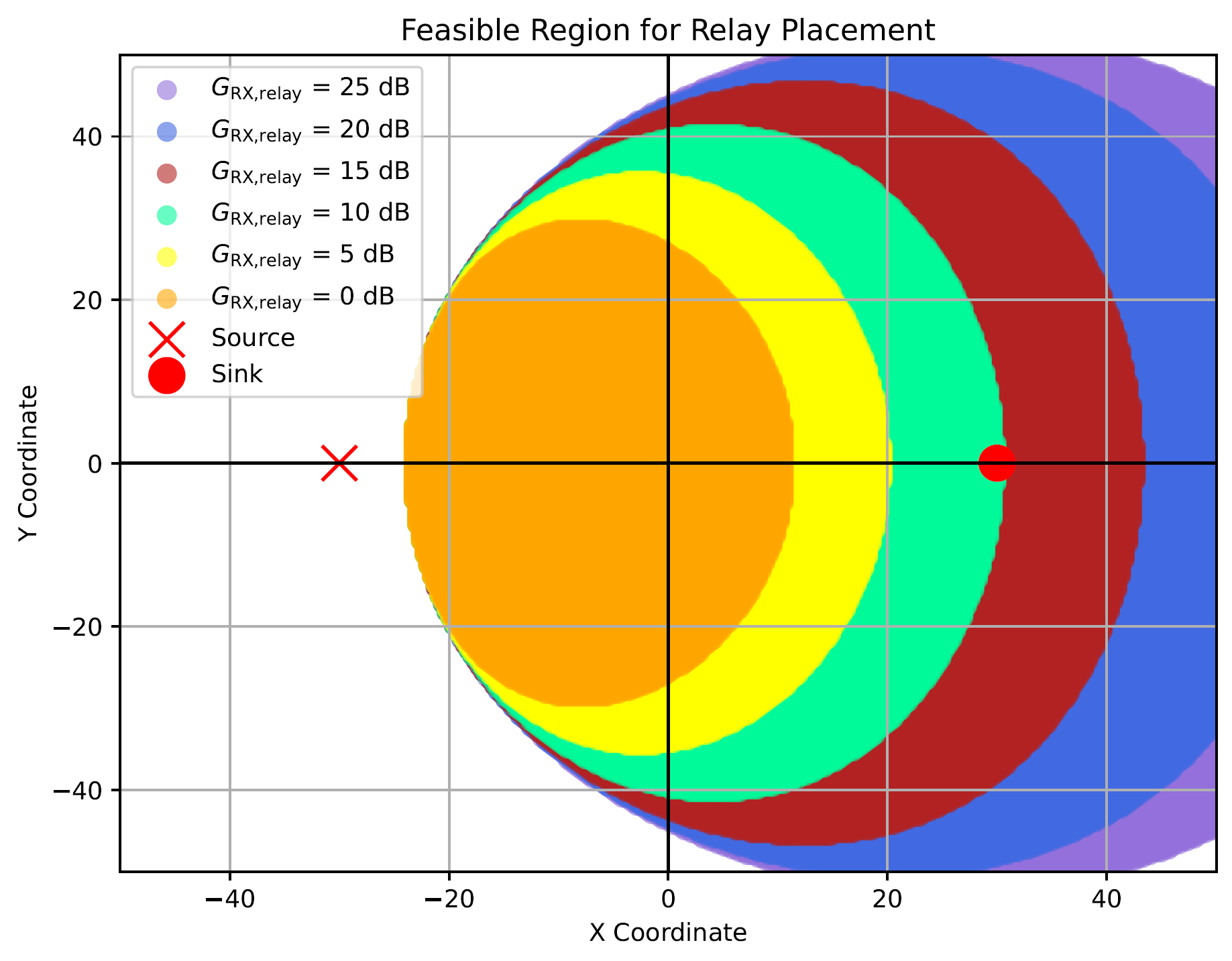}
    \caption{The advantageous regions to place the relay are given in this figure when the relay receiver gain is varied. In this setup, the source transmitter has a waste factor of unity, the relay transmitter has a waste factor of 2, the sink receiver has a 10 dB gain, and the path loss exponent is chosen to be 6. Every region includes the inner regions as well.}
    \label{g_vary}
\end{figure}

\subsection{Case Study: Fixed Wireless Access}

The adoption of FWA is accelerating worldwide, driven by advancements in 5G technology, increasing demand for broadband connectivity, and strategic investments by mobile service providers.
According to Ericsson’s Mobility Report \cite{ericsson2024mobility}, 79\% of mobile service providers now offer FWA, with 54\% providing it over 5G. FWA has become a key alternative to traditional fixed broadband, offering a cost-effective solution in regions where deploying fiber or cable is not feasible. As a result, global FWA connections are expected to more than double, from 160 million in 2024 to 350 million by 2030, with nearly 80\% of these operating on 5G networks.

FWA is also playing a growing role in mobile network traffic, with data consumption expected to increase more than fourfold by 2030, accounting for 36\% of total global mobile data traffic. Regional markets are experiencing rapid growth, with 5G FWA connections in India reaching nearly 3 million within just a year of launch, and Europe leading 73\% of all new 5G FWA deployments. The expansion of FWA is further supported by shifting business models, with 43\% of service providers now offering speed-based tariff plans, making FWA a competitive broadband alternative. Additionally, technological innovations such as 5G standalone (SA) deployments and more efficient customer premises equipment (CPE) are enhancing both performance and affordability \cite{ericsson2024mobility}.

\begin{figure}
    \centering
    \includegraphics[width=0.8\linewidth]{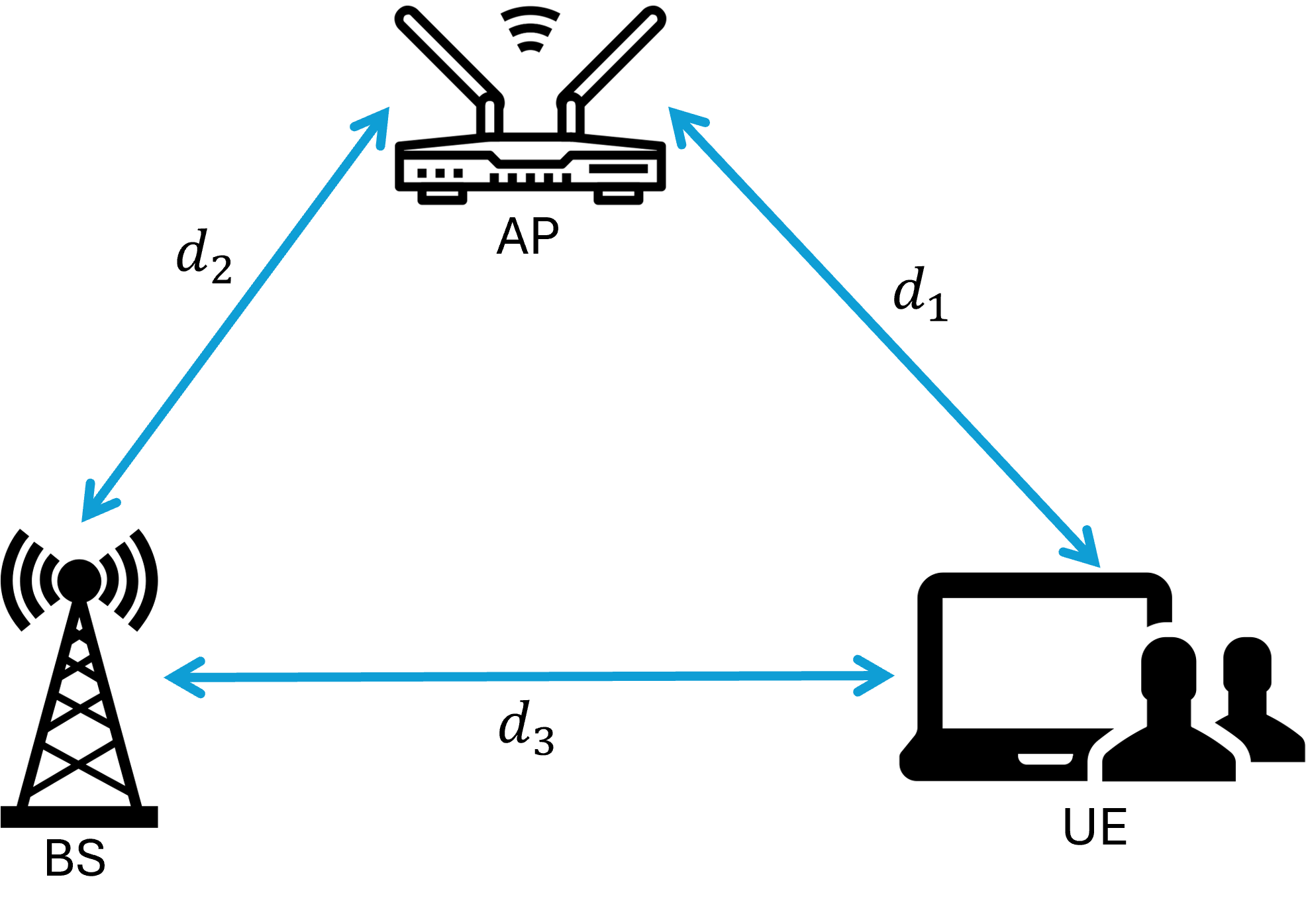}
    \caption{A general Fixed Wireless Access (FWA) system. We consider the links in both directions, i.e. uplink and downlink. During downlink transmission, the base station acts as the source, and the user equipment acts as the sink while during uplink transmission the roles are reversed. The access point acts as the relay in both cases.}
    \label{fwa-system}
\end{figure}

With FWA’s growing prominence, deciding on whether it is more advantageous to use an access point instead of a direct link is critical to maximizing energy efficiency and network performance. Poorly chosen links can lead to higher power consumption and reduced quality of service, impacting both operators and users. By leveraging our previously derived equations and decision rules for relay networks, we can analyze FWA in an energy consumption context. We can also see the effect of energy efficiency at user equipment and base station on the design of the system.

Let us now reinterpret the scenario shown in Figure 3 within the context of an FWA system, illustrated in Figure 8. In this setup, the key components are the user equipment (UE), an access point (AP), and a base station (BS). For this case study, we analyze both uplink and downlink communication paths and begin by calculating the minimum energy required to transmit a single bit over each link.

Starting with the direct uplink, where the UE transmits directly to the BS, the minimum energy per bit can be written as:
\begin{equation}
    E_{3}^{u} = 
    \frac{P_{NP_3}^{u}}{C_{3}^{u}} + 
    N_0 \ln(2)\, W_{3}^{u},
    \label{start}
\end{equation}
where \( P_{NP_3}^{u} \) denotes the non-path power consumption, \( C_{3}^{u} \) is the uplink capacity, and \( W_{3}^{u} \) is the waste factor associated with the uplink.

The uplink waste factor can be approximated using:
\begin{equation}
    W_{3}^{u} \approx \frac{W_{\text{TX,UE}}}{G_{\text{RX,BS}} G_{3}},
\end{equation}
where \( W_{\text{TX,UE}} \) is the waste factor of the UE’s transmitter, \( G_{\text{RX,BS}} \) is the gain of the BS receiver, and \( G_{3} \) is the channel gain, which remains the same for both uplink and downlink.

Similarly, the energy consumption per bit in the downlink, where the BS transmits to the UE, is given by:
\begin{equation}
    E_{3}^{d} = 
    \frac{P_{NP_3}^{d}}{C_{3}^{d}} + 
    N_0 \ln(2)\, W_{3}^{d}.
\end{equation}

The corresponding downlink waste factor is approximated as:
\begin{equation}
    W_{3}^{d} \approx \frac{W_{\text{TX,BS}}}{G_{\text{RX,UE}} G_{3}},
    \label{end}
\end{equation}
where \( W_{\text{TX,BS}} \) is the BS transmitter’s waste factor and \( G_{\text{RX,UE}} \) is the receiver gain at the UE.

To evaluate the total energy expenditure over a bidirectional wireless link, it is natural to consider both uplink and downlink contributions. One intuitive approach is to compute a weighted sum of energy expenditures based on the proportion of traffic flowing in each direction. Let \( \rho_{\text{d}} \) and \( \rho_{\text{u}} \) denote the fractions of total traffic in the downlink and uplink, respectively, where \( \rho_{\text{d}} + \rho_{\text{u}} = 1 \). The combined energy consumption per bit, \( E_3 \), over the direct link can then be expressed as:
\begin{equation}
    E_3 = \rho_{\text{d}} E_3^{\text{d}} + \rho_{\text{u}} E_3^{\text{u}}
\end{equation}

Substituting from Eqs.~\ref{start}–\ref{end}, we obtain:

\begin{equation}
    \begin{split}
        E_3 = \rho_{\text{d}} \left( \frac{P_{NP_3}^{\text{d}}}{C_3^{\text{d}}} + 
        N_0 \ln(2) W_3^{\text{d}} \right)
        + \rho_{\text{u}} \left( \frac{P_{NP_3}^{\text{u}}}{C_3^{\text{u}}} + 
        N_0 \ln(2) W_3^{\text{u}} \right)
    \end{split}
\end{equation}

By rearranging terms, the expression becomes:

\begin{equation}
    \begin{split}
        E_3 = \left( 
            \rho_{\text{d}} \frac{P_{NP_3}^{\text{d}}}{C_3^{\text{d}}} + 
            \rho_{\text{u}} \frac{P_{NP_3}^{\text{u}}}{C_3^{\text{u}}}
        \right) 
        + N_0 \ln(2) \left( 
            \rho_{\text{d}} W_3^{\text{d}} + 
            \rho_{\text{u}} W_3^{\text{u}} 
        \right)
    \end{split}
\end{equation}

Applying the same methodology to the link that routes through the access point yields the total energy per bit, \( E_{12} \), as follows (intermediate steps are omitted for brevity):

\begin{equation}
    \begin{split}
        E_{12} = 
        \left( 
            \rho_{\text{u}} \left( 
                \frac{P_{NP_1}^{\text{u}}}{C_1^{\text{u}}} + 
                \frac{P_{NP_2}^{\text{u}}}{C_2^{\text{u}}}
            \right) \right. 
        \left. + 
            \rho_{\text{d}} \left( 
                \frac{P_{NP_1}^{\text{d}}}{C_1^{\text{d}}} + 
                \frac{P_{NP_2}^{\text{d}}}{C_2^{\text{d}}}
            \right)
        \right) \\
        + N_0 \ln(2) \left( 
            \rho_{\text{u}} (W_1^{\text{u}} + W_2^{\text{u}}) +
            \rho_{\text{d}} (W_1^{\text{d}} + W_2^{\text{d}})
        \right)
    \end{split}
\end{equation}

\begin{figure}[h]
    \centering
    \includegraphics[width=0.9\linewidth]{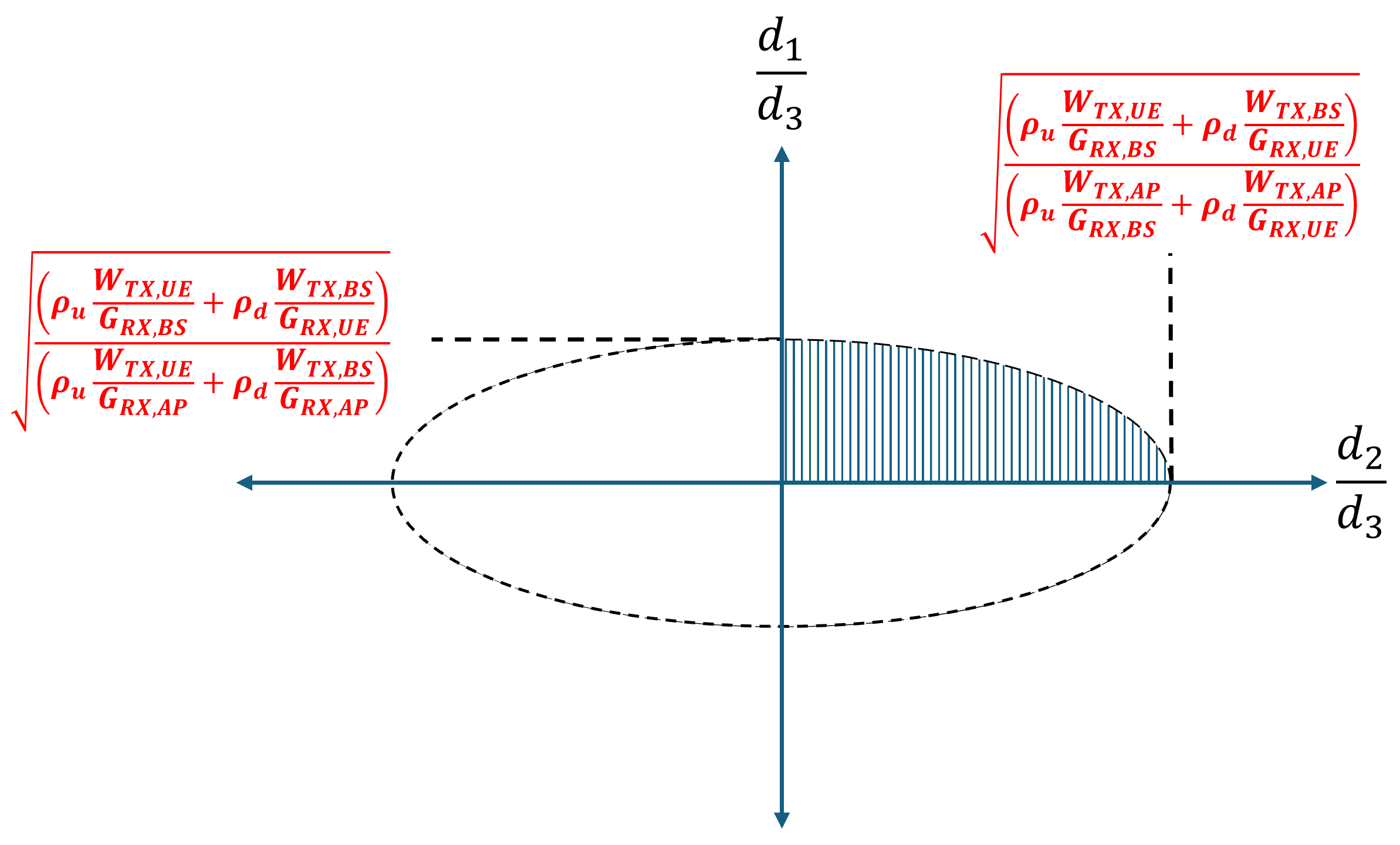}
    \caption{Based on Eq. \ref{distance-3}, the striped area denotes the region where it is advantageous to use a relay. In this figure, the path loss exponent is assumed to be 2, the non-path power consumption is assumed to be 0, and $G_{RX}G_{ch}<<1$.}
    \label{ellipse-2}
\end{figure}

The waste factors in the respective directions are given by:
{\small
\begin{equation}
    \begin{split}
        W_1^{\text{u}} \approx \frac{W_{\text{TX,UE}}}{G_{\text{RX,AP}} G_1}, \hfill 
        W_2^{\text{u}} \approx \frac{W_{\text{TX,AP}}}{G_{\text{RX,BS}} G_2} \\
        W_1^{\text{d}} \approx \frac{W_{\text{TX,BS}}}{G_{\text{RX,AP}} G_1}, \hfill 
        W_2^{\text{d}} \approx \frac{W_{\text{TX,AP}}}{G_{\text{RX,UE}} G_2}
    \end{split}
\end{equation}}

Assuming that the non-path power consumption and the channel capacities are equal across all links, we denote them as \( P_{NP} \) and \( C \), respectively. Under this assumption, we can express the ratio of the total energy consumed per bit using the access point, \( E_{12} \), to that of the direct link, \( E_3 \), as in Eq.~\eqref{ratio}. Note that we, again, require \( E_{12}/E_3 \) ratio to be less than 1 to ensure the direct link is more costly.

    \begin{equation}
    \frac{E_{12}}{E_3} =
    \frac{
        2 \frac{P_{NP}}{C} (\rho_{\text{u}} + \rho_{\text{d}}) + 
        N_0 \ln(2) \left( 
            \rho_{\text{u}} (W_1^{\text{u}} + W_2^{\text{u}}) + 
            \rho_{\text{d}} (W_1^{\text{d}} + W_2^{\text{d}})
        \right)
    }{
        \frac{P_{NP}}{C} (\rho_{\text{u}} + \rho_{\text{d}}) + 
        N_0 \ln(2) \left( 
            \rho_{\text{d}} W_3^{\text{d}} + 
            \rho_{\text{u}} W_3^{\text{u}} 
        \right)
    } < 1
\label{ratio}
\end{equation}

To determine when using the access point becomes advantageous over the direct link, we follow a similar derivation approach as in Eq.~\ref{distance-2}. Skipping intermediate algebraic steps for brevity, the resulting inequality yields a geometric condition based on the distances between nodes and system parameters:

\begin{equation}
\begin{split}
    d_3^\alpha >\ 
    &d_1^\alpha 
    \frac{
        \rho_{\text{u}} \frac{W_{\text{TX,UE}}}{G_{\text{RX,AP}}} + 
        \rho_{\text{d}} \frac{W_{\text{TX,BS}}}{G_{\text{RX,AP}}}
    }{
        \rho_{\text{u}} \frac{W_{\text{TX,UE}}}{G_{\text{RX,BS}}} + 
        \rho_{\text{d}} \frac{W_{\text{TX,BS}}}{G_{\text{RX,UE}}}
    } + d_2^\alpha 
    \frac{
        \rho_{\text{u}} \frac{W_{\text{TX,AP}}}{G_{\text{RX,BS}}} + 
        \rho_{\text{d}} \frac{W_{\text{TX,AP}}}{G_{\text{RX,UE}}}
    }{
        \rho_{\text{u}} \frac{W_{\text{TX,UE}}}{G_{\text{RX,BS}}} + 
        \rho_{\text{d}} \frac{W_{\text{TX,BS}}}{G_{\text{RX,UE}}}
    }
\end{split}
\label{fwa-distance}
\end{equation}

The full derivation for Eq.~\eqref{fwa-distance} is provided in the Supplementary Information section at the end of this paper.

The expression in Eq.~\eqref{fwa-distance} gives an explicit condition on the maximum distance \( d_3 \) (between the user equipment and the base station) for which using an access point becomes more energy-efficient. The expression incorporates both directional traffic proportions and the efficiency and gain characteristics of all involved components.

To gain further geometric intuition, again, we consider the free-space loss model, where $\alpha=2$. We obtain a similar ellipses condition as in Section IV by arranging some terms:

\begin{equation}
    1 > \frac{\left(
        \frac{d_1}{d_3}
    \right)^2}{\left(\frac{\rho_{\text{u}} \frac{W_{\text{TX,UE}}}{G_{\text{RX,BS}}} + 
        \rho_{\text{d}} \frac{W_{\text{TX,BS}}}{G_{\text{RX,UE}}}}{\rho_{\text{u}} \frac{W_{\text{TX,UE}}}{G_{\text{RX,AP}}} + 
        \rho_{\text{d}} \frac{W_{\text{TX,BS}}}{G_{\text{RX,AP}}}}\right)}
     + \frac{\left(
        \frac{d_2}{d_3}
    \right)^2}{\left( \frac{\rho_{\text{u}} \frac{W_{\text{TX,UE}}}{G_{\text{RX,BS}}} + 
        \rho_{\text{d}} \frac{W_{\text{TX,BS}}}{G_{\text{RX,UE}}}}{\rho_{\text{u}} \frac{W_{\text{TX,AP}}}{G_{\text{RX,BS}}} + 
        \rho_{\text{d}} \frac{W_{\text{TX,AP}}}{G_{\text{RX,UE}}}} \right)}
        \label{fwa-dist}
\end{equation}

The condition is visualized in Figure 9 as an ellipse condition. Since we still have non-negativity conditions on distance ratios, only the first quadrant of the ellipse is valid.

{
To further illustrate the implications of Eqs.}~\eqref{fwa-distance} { and} \eqref{fwa-dist}, {consider two extreme traffic scenarios:}

First, in an \textit{uplink-heavy traffic} case where \( \rho_u \gg \rho_d \), the terms multiplied by \( \rho_u \) dominate the inequality. In this regime, Eq.~\eqref{fwa-dist} simplifies to:
\begin{equation}
    1 > \frac{\left(
        \frac{d_1}{d_3}
    \right)^2}{\left(\frac{G_{\text{RX,AP}}}{G_{\text{RX,BS}}}\right)}
     + \frac{\left(
        \frac{d_2}{d_3}
    \right)^2}{\left( \frac{W_{\text{TX,UE}}}{W_{\text{TX,AP}}} \right)}.
    \label{edge-up}
\end{equation}

Second, in a \textit{downlink-heavy traffic} case where \( \rho_u \ll \rho_d \), the terms multiplied by \( \rho_d \) dominate, and Eq.~\eqref{fwa-dist} reduces to:
\begin{equation}
    1 > \frac{\left(
        \frac{d_1}{d_3}
    \right)^2}{\left(\frac{G_{\text{RX,AP}}}{G_{\text{RX,UE}}}\right)}
     + \frac{\left(
        \frac{d_2}{d_3}
    \right)^2}{\left( \frac{W_{\text{TX,BS}}}{W_{\text{TX,AP}}} \right)}.
    \label{edge-down}
\end{equation}

These two forms are structurally equivalent to Eq.~\eqref{distance-3}, where the AP acts as a relay between the UE and the BS. In the uplink-dominant case, the UE predominantly acts as the source and the BS as the sink. Conversely, in the downlink-dominant case, these roles are reversed.

Therefore, Eqs.~\eqref{fwa-distance} and \eqref{fwa-dist} can be viewed as generalized versions of Eqs.~\eqref{distance-2} and \eqref{distance-3}, accommodating asymmetric traffic conditions. This generalization highlights the flexibility of the waste factor framework in adapting to varying traffic loads.

\begin{figure}
    \centering
    \includegraphics[width=\linewidth]{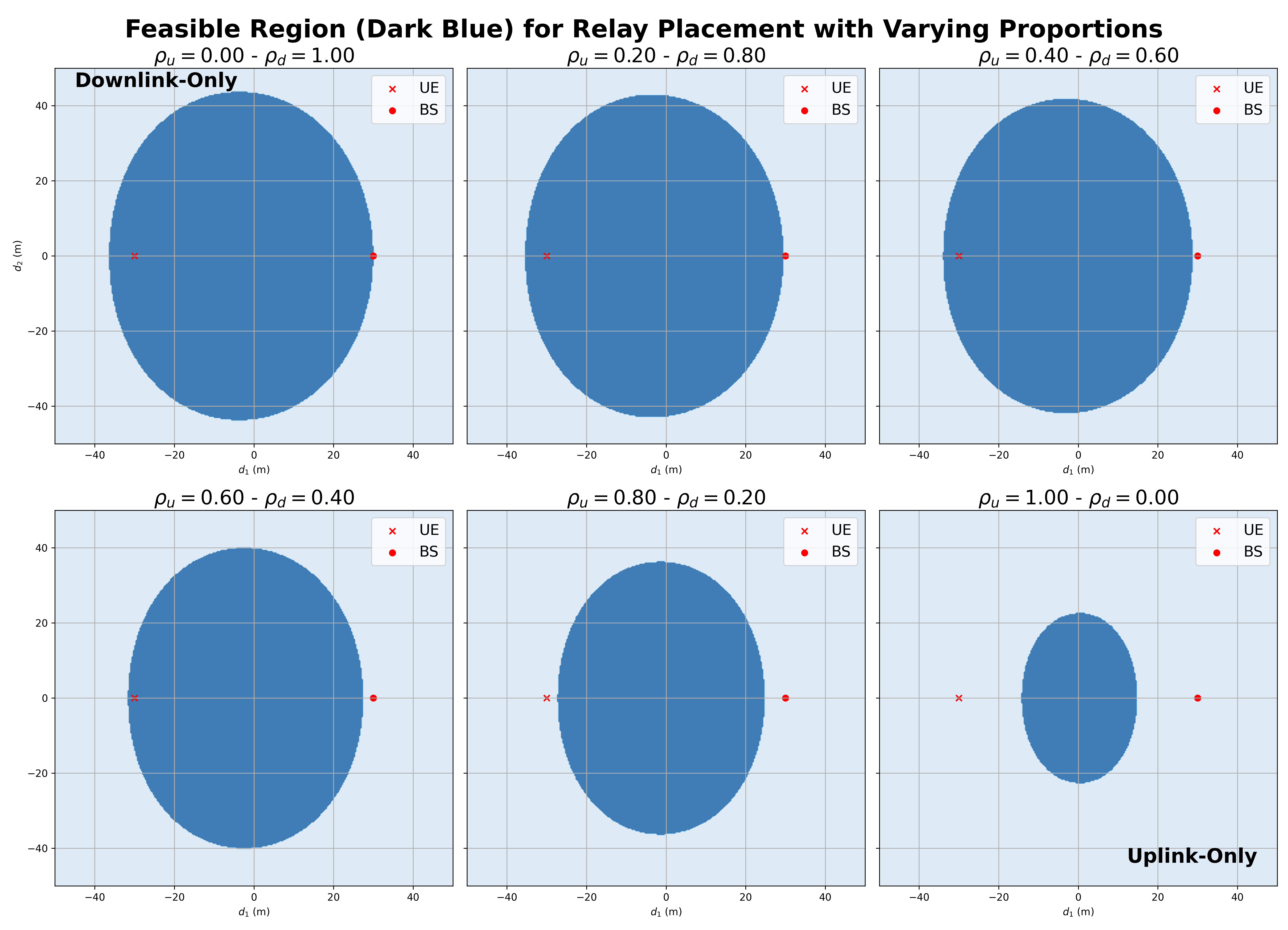}
    \caption{The advantageous regions to place the relay are given in this figures with dark blue color when the traffic proportions are varied. In this setup, $W_{\text{TX,BS}}=15$, $W_{\text{TX,UE}}=3$, $W_{\text{TX,AP}}=10$, $G_{\text{RX,BS}}=15 \text{dB}$, $G_{\text{RX,UE}}=10 \text{dB}$, $G_{\text{RX,AP}}=10 \text{dB}$, and the path loss exponent $\alpha$ is chosen to be 4. The dark blue areas indicate the feasible regions.}
    \label{fig:enter-label}
\end{figure}

{To supplement the analytical framework with a more realistic scenario, Figure} 10 {illustrates how the advantageous relay region evolves as the uplink/downlink traffic proportions are varied. In this setup, the BS transmitter has a waste factor of 15, the UE transmitter a waste factor of 3, and the AP transmitter a waste factor of 10. Receiver gains are set to 15 dB for the BS, 10 dB for both the UE and the AP, and the path loss exponent is taken as 4, representing an urban fixed wireless access (FWA) environment. The dark blue area in each plot denotes the region where relaying is more advantageous than direct transmission.}

In accordance with Eq.~\eqref{edge-up}, we observe that the feasible relay region contracts under uplink-heavy traffic and expands under downlink-heavy conditions. This behavior is due to the asymmetry in system parameters: the BS receiver gain (\( G_{\text{RX,BS}} \)) is higher than that of the UE, and the UE transmitter has significantly lower waste (\( W_{\text{TX,UE}} \)) compared to the BS transmitter. These trends validate the theoretical conditions derived in Eqs.~\eqref{fwa-distance}–\eqref{edge-down} under practical assumptions and demonstrate how traffic asymmetry and hardware characteristics jointly affect relay deployment decisions.

\color{black}

\subsection{Analogy With Friis' Noise Figure Formula: A Discussion} 

In this paper, we use the concept of the waste factor $W$ in analogy with Friis' formula for the noise figure of cascaded elements in a receiver chain. In this section, we extend the discussion to gain additional insights into cascaded systems through the lens of $W$. While the analyses in Sections IV and V established relationships between hop distances and system performance, a deeper exploration of the system structure reveals additional considerations that can guide the design of multi-hop networks. 
For reference, Eq.~\eqref{eq:Wchain} is repeated below:
\begin{equation*}
    \begin{split}
        W &= \left\{  W_N + \frac{W_{N-1} - 1}{G_N} + \frac{W_{N-2} - 1}{G_N G_{N-1}} + \cdots + \frac{W_1 - 1}{G_N \dots G_2} \right\}.
    \end{split}
\end{equation*}

To reveal finer structure, Figure 11 (top) shows a more detailed breakdown of the transmission chain, proceeding from sink to source: low-noise amplifier (LNA), receiver antenna, channel, transmitter antenna, power amplifier (PA), and processing block. The waste factor expression for the setup in Figure 11 can be written as:
\begin{equation}
    \begin{split}
        W &= \left\{  W_N^{(lna)} + \frac{1/G_{N-1}^{(rx)} - 1}{G_N^{(lna)}} + \frac{1/G_{N-2}^{(ch)} - 1}{G_N^{(lna)} G_{N-1}^{(rx)}}  + \right. \\
        &\frac{1/G_{N-3}^{(tx)} - 1}{G_N^{(lna)} G_{N-1}^{(rx)} G_{N-2}^{(ch)}} + \frac{W_{N-4}^{(pa)} - 1}{G_N^{(lna)} G_{N-1}^{(rx)} G_{N-2}^{(ch)} G_{N-3}^{(tx)}} \\
        &\left. + \frac{W_{N-5}^{(proc)} - 1}{G_N^{(lna)} \cdots G_{N-4}^{(pa)}} + \frac{W_{N-6}^{(rx)} - 1}{G_N^{(lna)} \cdots G_{N-5}^{(proc)}} \right\}.
    \end{split}
    \label{eq:detailedW}
\end{equation}

\begin{figure}[H]
    \centering
        \includegraphics[width=\textwidth]{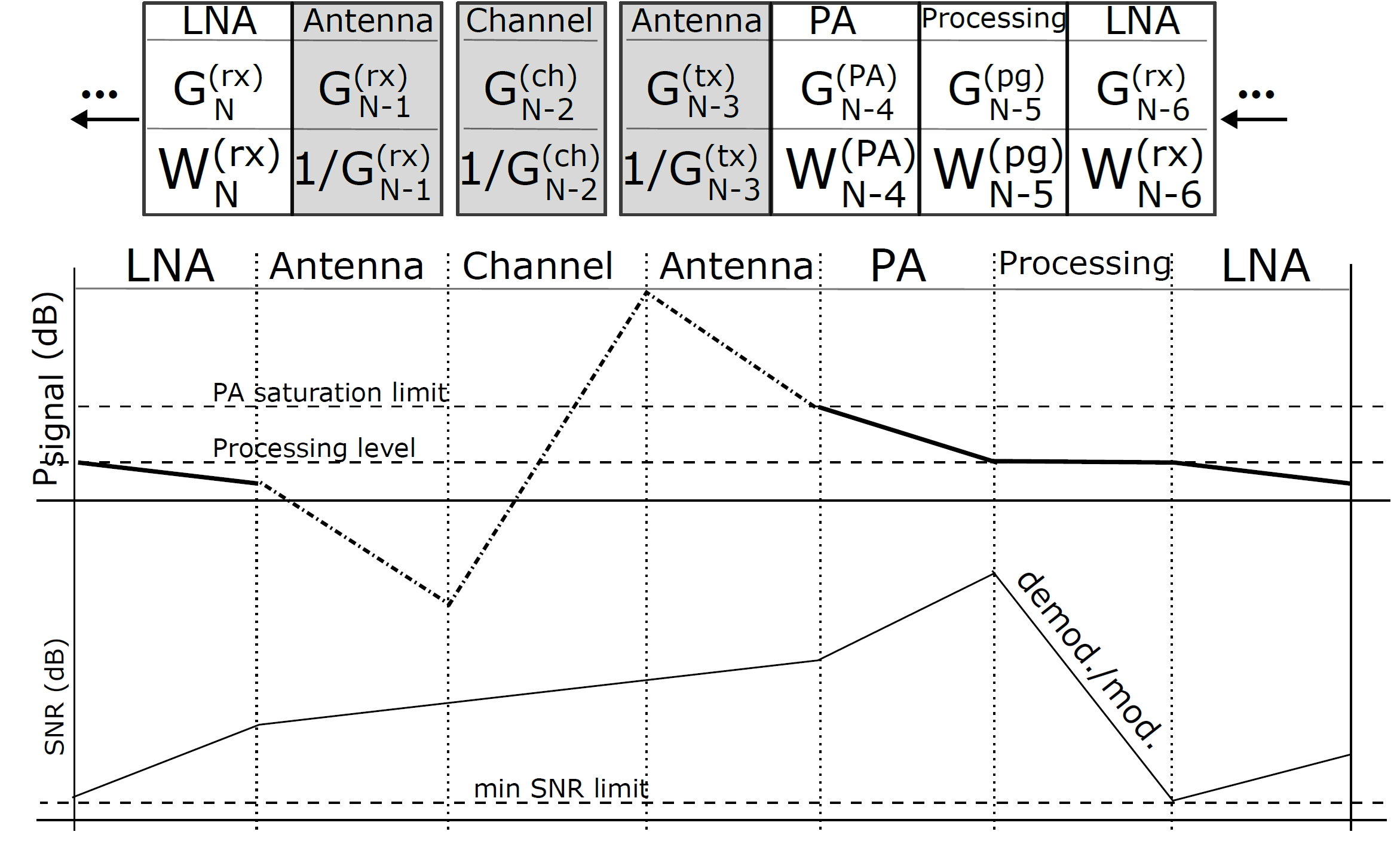}
        \label{fig:detailedchain-1}
    \caption{\textbf{(top)} Granular stage breakdown from $N^\text{th}$ to $(N-6)^\text{th}$ element. \textbf{(bottom)} Power and SNR evolution across the stages.}
    \label{fig:detailedchain}
\end{figure}

For comparison, Friis' equation for the total noise factor $F_\text{total}$ across cascaded stages is given by:
\begin{equation}
    F_\text{total} = F_1 + \frac{F_2 - 1}{G_1} + \frac{F_3 - 1}{G_1 G_2} + \cdots.
\end{equation}

While the mathematical structures are similar, dissecting the system highlights key conceptual differences:
\begin{itemize}
    \item In Friis' formula, the first-stage noise figure $F_1$ is critical because the signal is weakest at the input. Subsequent stages have diminishing influence.
    \item In contrast, in the waste factor formulation, the dominant term is not necessarily $W_1$ or $W_N$. Instead, the significance of each term depends on the signal energy at that stage.
\end{itemize}

Specifically, stages that handle higher energy levels have a greater influence on $W$, which becomes clear in Eq.~\eqref{eq:detailedW}: the dominant term is associated with the power amplifier stage:
\begin{equation}
\frac{W_{N-4}^{(pa)} - 1}{G_N^{(lna)} G_{N-1}^{(rx)} G_{N-2}^{(ch)} G_{N-3}^{(tx)}}
\label{dominant-term}
\end{equation}
The term in \eqref{dominant-term} is critical because:
\begin{itemize}
    \item It precedes the air-link stages, where losses are high (i.e., $G_{N-2}^{(ch)}$ is small).
    \item Its denominator includes the channel loss, which amplifies its impact.
    \item It is followed by stages (transmit antenna, channel, receive antenna, LNA) whose gains collectively shape its influence.
\end{itemize}
Thus, antenna design and link budgeting can play critical roles in shaping $W$. Careful design can mitigate waste accumulation, a realization that provides finer granularity beyond the approximations used earlier.

Another important insight is that, unlike Friis' cascaded chain where signal levels generally increase, the waste factor chain exhibits \textit{cyclic behavior} in energy and SNR. As we move across stages:
\begin{itemize}
    \item The energy and SNR form cycles between stages (Figure 11 (bottom)).
    \item Transmission energy peaks at the PA output (limited by saturation).
    \item Energy falls across the antennas and channel (due to spreading loss and noise).
    \item SNR is lowest at the receiver LNA input and highest after processing.
\end{itemize}

{The granular breakdown in Figure 11 (top) and the cyclic energy/SNR behavior in Figure 11 (bottom) highlight that stages handling higher absolute energy levels—particularly the PA and beamforming network—dominate the end-to-end waste factor. This suggests a hardware design trade‑off: investing in higher PA efficiency or lower-loss beamforming architectures can yield outsized reductions in $W$, compared to marginal improvements in other low‑power stages. Designers can, therefore, use the \(W\) partition to prioritize component R\&D where it matters most.}
{
The modular ``energy cycle'' view of each hop implies that multi-hop links can be analyzed and optimized on a per-hop basis, rather than treated as a monolithic cascade as in Friis' noise figure analysis. Each relay hop exhibits its own PA--antenna--channel--LNA cycle; this decoupling opens new design possibilities for scalable, high-efficiency mesh and ad hoc networks. While equalizing the marginal waste contributions across hops (i.e., setting }$\Delta W~ ${ equal per hop) may offer design simplicity or fairness, we leave formal optimization of waste distribution for future work. Nevertheless, it should be clear from Eq.}~\eqref{waste-factor-2}{ that the analysis of various cascades of components may be cast in a simple mathematical framework suitable for analysis, measurement, and optimization of the various components along a cascaded system.}

In dynamic or mobile scenarios, one can adapt operating points—such as beam‑width, PA back‑off, or relay activation—to minimize instantaneous $W$. For example, in light obstruction conditions mmWave arrays can narrow beams to reduce waste, while in heavy blockage they can switch to wider beams or sub‑6 GHz fallback to trade spectral efficiency for energy savings. Such adaptive control policies could be derived directly from real‑time estimates of per‑stage $W$.

{Finally, $W$ captures absolute power waste, it can inform regulatory and environmental metrics. Network operators and standards bodies (e.g. ETSI, 3GPP) could employ $W$-based thresholds to limit out‑of‑band emissions or to certify “green” base stations based on end‑to‑end energy efficiency rather than component‑level specifications alone. }

\section{Discussion}

This paper has introduced the \textit{Waste Factor} (W) as a rigorous and physically grounded metric for quantifying energy efficiency in wireless communication systems. Drawing inspiration from Friis’ classical noise factor and motivated by the urgent need to understand and reduce energy expenditure in next-generation networks, the Waste Factor offers a unified framework to assess how power is utilized and dissipated across cascaded systems. Unlike traditional efficiency metrics that often obscure internal energy flows, W captures both the signal-contributing and non-contributing power pathways, providing system designers with transparent and actionable insights.

We began by formulating the Waste Factor at the component level and generalized it to arbitrary cascaded systems, highlighting its gain-aware, recursive structure. This formulation enables intuitive and tractable expressions for the total system waste, which we demonstrated for multi-stage links and relay-based architectures. By introducing the channel as a cascade node with known attenuation properties, we extended the scope of W to encompass full wireless transmission paths.

To relate this framework to theoretical bounds, we revisited the \textit{Consumption Factor} (CF) originally introduced in prior literature as a unifying performance metric. We re-derived CF in terms of the Waste Factor, showing how energy efficiency limits—traditionally grounded in Shannon’s minimum energy per bit—can be augmented to include additive signal waste along the cascade. This is a novel contribution that, for the first time, embeds W directly into the Shannon limit formulation. In the wideband regime, we derived the limiting behavior of CF and obtained a closed-form expression for the minimum bit energy $E_{bc}$ consumed by a system, which includes both non-path power and the waste-weighted signal energy.

Building on this foundation, we derived analytical conditions for when a relay is more energy-efficient than a direct wireless link, casting the results as inequalities in terms of component gains, path loss exponents, and waste factors. These results were visualized as elliptical relay regions under various system parameters, providing geometric and design intuition for practical deployment scenarios. Notably, since a RIS can be interpreted as a passive relay element, the derived conditions and visualizations also offer useful insight for RIS-assisted system design and placement.

In the case of passive RIS, the waste factor framework aligns naturally with modeling the RIS as a purely passive, loss-only stage, consistent with the formulation in Eq.~\eqref{channel-wf}. However, emerging active RIS architectures—equipped with amplifiers, active load networks, or reconfigurable electronics—break this passive assumption. These systems introduce non-negligible power consumption at the RIS node itself and may also add amplification noise, requiring both path-related and non-path-related power components to be explicitly modeled. Extending the waste factor framework to active RIS systems involves treating these local power-consuming elements as part of the signal path cascade between source and sink. Such an approach has been recently explored for MIMO systems \cite{ying_waste_globecom_24}, where active and passive components are jointly modeled to optimize energy efficiency using the Waste Figure. Applying this framework to RIS-assisted systems represents a promising direction for future work, enabling more accurate energy-aware RIS deployment and configuration strategies.

The analysis was further extended to Fixed Wireless Access (FWA) systems, where asymmetric traffic in uplink and downlink directions was explicitly accounted for. We derived a generalized energy consumption rule that combines link-level waste with traffic weighting, enabling infrastructure-level decisions such as whether to route data through an access point or directly to the base station. This use case exemplifies the practical relevance of Waste Factor in guiding efficient wireless network topologies.

Ultimately, this work offers a comprehensive, extensible, and physically meaningful framework for energy-aware system design in wireless communications. As the energy demands of the ICT industry accelerate—particularly with the scaling of AI and cloud-based services—the Waste Factor serves as a critical bridge between device-level modeling and network-wide energy optimization. 

{While the Waste Factor} $W${ is defined analytically in terms of signal and power flows across RF and digital components, practical measurement or monitoring in deployed networks are easily conducted as described in} \cite{rappaport24waste,ying2023waste}{. All that is needed is system-level instrumentation that measures a) output signal power, and b) total consumed power from the system power supply. In a base station or relay node,} $W${ could be monitored by tracking the ratio of power delivered to the antenna (or subsequent stage) versus the total power drawn by the stage. This can be achieved using existing power monitoring modules at the amplifier output, power supply telemetry, and digital baseband counters tracking signal throughput. Additionally, known overheads such as pilot insertion or control signaling (contributing to} $P_{NS}$) {can be inferred from protocol layer counters. Control-plane signaling could then periodically report these metrics (e.g., every few seconds or based on network load) to a centralized optimizer. While the frequency of updates may depend on system dynamics, we anticipate that slow-timescale monitoring (on the order of seconds) is sufficient for energy-aware scheduling decisions. The overhead of this reporting is minimal, especially in 5G/6G base stations where power telemetry and performance counters are already integral for health monitoring and scheduling. In fact, we believe that most present installations of wireless infrastructure could support the on-going measurement of Waste Figure with minimal to no retrofitting.}

Other future research directions include extending this framework to multi-user and multi-hop cases, analyzing the asymptotic behavior of the formulation. Moreover, integration with learning-based and optimization-driven network control schemes may enable real-time adaptation of transmission strategies using W and CF as embedded objectives. The integration of AI may also help develop strategies when the system parameters and element gains are not fixed. In such adaptive systems, where transmit power, modulation, and coding schemes vary over time, AI-driven models can learn to estimate or predict waste factors dynamically based on system telemetry, channel conditions, or traffic patterns. This enables real-time energy-aware decisions such as link adaptation, relay selection, or beam configuration, while accounting for both efficiency and performance.

In the zettajoule era, metrics that unify theoretical limits with physical energy realities—like Waste Factor—will be essential to sustainable wireless evolution.

\section{Data Availability}
Data sharing is not applicable to this article as no datasets were generated or analysed during the current study.

\section{Acknowledgements}
This material is based upon work supported in part by the U.S. Department of Energy, Office of Science, Office of Advanced Scientific Computing Research, Early Career Research Program under Award Number DE-SC0023957, and in part by the Hagler Institute for Advanced Study at Texas A\&M University.

\section{Author contributions}
All authors conceived the idea, performed the research, and wrote the
manuscript.

\section{Competing interests}
The authors declare no competing interests.


\bibliography{references}


\begin{thebibliography}{37}
\ifx \bisbn   \undefined \def \bisbn  #1{ISBN #1}\fi
\ifx \binits  \undefined \def \binits#1{#1}\fi
\ifx \bauthor  \undefined \def \bauthor#1{#1}\fi
\ifx \batitle  \undefined \def \batitle#1{#1}\fi
\ifx \bjtitle  \undefined \def \bjtitle#1{#1}\fi
\ifx \bvolume  \undefined \def \bvolume#1{\textbf{#1}}\fi
\ifx \byear  \undefined \def \byear#1{#1}\fi
\ifx \bissue  \undefined \def \bissue#1{#1}\fi
\ifx \bfpage  \undefined \def \bfpage#1{#1}\fi
\ifx \blpage  \undefined \def \blpage #1{#1}\fi
\ifx \burl  \undefined \def \burl#1{\textsf{#1}}\fi
\ifx \doiurl  \undefined \def \doiurl#1{\url{https://doi.org/#1}}\fi
\ifx \betal  \undefined \def \betal{\textit{et al.}}\fi
\ifx \binstitute  \undefined \def \binstitute#1{#1}\fi
\ifx \binstitutionaled  \undefined \def \binstitutionaled#1{#1}\fi
\ifx \bctitle  \undefined \def \bctitle#1{#1}\fi
\ifx \beditor  \undefined \def \beditor#1{#1}\fi
\ifx \bpublisher  \undefined \def \bpublisher#1{#1}\fi
\ifx \bbtitle  \undefined \def \bbtitle#1{#1}\fi
\ifx \bedition  \undefined \def \bedition#1{#1}\fi
\ifx \bseriesno  \undefined \def \bseriesno#1{#1}\fi
\ifx \blocation  \undefined \def \blocation#1{#1}\fi
\ifx \bsertitle  \undefined \def \bsertitle#1{#1}\fi
\ifx \bsnm \undefined \def \bsnm#1{#1}\fi
\ifx \bsuffix \undefined \def \bsuffix#1{#1}\fi
\ifx \bparticle \undefined \def \bparticle#1{#1}\fi
\ifx \barticle \undefined \def \barticle#1{#1}\fi
\bibcommenthead
\ifx \bconfdate \undefined \def \bconfdate #1{#1}\fi
\ifx \botherref \undefined \def \botherref #1{#1}\fi
\ifx \url \undefined \def \url#1{\textsf{#1}}\fi
\ifx \bchapter \undefined \def \bchapter#1{#1}\fi
\ifx \bbook \undefined \def \bbook#1{#1}\fi
\ifx \bcomment \undefined \def \bcomment#1{#1}\fi
\ifx \oauthor \undefined \def \oauthor#1{#1}\fi
\ifx \citeauthoryear \undefined \def \citeauthoryear#1{#1}\fi
\ifx \endbibitem  \undefined \def \endbibitem {}\fi
\ifx \bconflocation  \undefined \def \bconflocation#1{#1}\fi
\ifx \arxivurl  \undefined \def \arxivurl#1{\textsf{#1}}\fi
\csname PreBibitemsHook\endcsname

\bibitem[\protect\citeauthoryear{{International Energy Agency}}{2025a}]{iea2025global}
\begin{botherref}
\oauthor{\bsnm{{International Energy Agency}}}:
Global Energy Review 2025.
Licence: CC BY 4.0.
\url{https://www.iea.org/reports/global-energy-review-2025}
\end{botherref}
\endbibitem

\bibitem[\protect\citeauthoryear{{International Energy Agency}}{2025b}]{iea2025energyai}
\begin{botherref}
\oauthor{\bsnm{{International Energy Agency}}}:
Energy and {AI}.
Technical report,
International Energy Agency
(2025).
Accessed: 2025-04-14.
\url{https://www.iea.org/reports/energy-and-ai}
\end{botherref}
\endbibitem

\bibitem[\protect\citeauthoryear{Lindsey}{2024}]{lindsey2024co2}
\begin{botherref}
\oauthor{\bsnm{Lindsey}, \binits{R.}}:
Climate Change: Atmospheric Carbon Dioxide.
Reviewed by Ed Dlugokencky. Published by NOAA Climate.gov.
\url{https://www.climate.gov/news-features/understanding-climate/climate-change-atmospheric-carbon-dioxide}
\end{botherref}
\endbibitem

\bibitem[\protect\citeauthoryear{Patterson et~al.}{2022}]{9810097}
\begin{barticle}
\bauthor{\bsnm{Patterson}, \binits{D.}},
\bauthor{\bsnm{Gonzalez}, \binits{J.}},
\bauthor{\bsnm{Hölzle}, \binits{U.}},
\bauthor{\bsnm{Le}, \binits{Q.}},
\bauthor{\bsnm{Liang}, \binits{C.}},
\bauthor{\bsnm{Munguia}, \binits{L.-M.}},
\bauthor{\bsnm{Rothchild}, \binits{D.}},
\bauthor{\bsnm{So}, \binits{D.R.}},
\bauthor{\bsnm{Texier}, \binits{M.}},
\bauthor{\bsnm{Dean}, \binits{J.}}:
\batitle{The carbon footprint of machine learning training will plateau, then shrink}.
\bjtitle{Computer}
\bvolume{55}(\bissue{7}),
\bfpage{18}--\blpage{28}
(\byear{2022})
\doiurl{10.1109/MC.2022.3148714}
\end{barticle}
\endbibitem

\bibitem[\protect\citeauthoryear{{International Energy Agency}}{2023}]{iea_data_centres}
\begin{botherref}
\oauthor{\bsnm{{International Energy Agency}}}:
Data Centres and Data Transmission Networks.
Accessed: 2025-04-13.
\url{https://www.iea.org/energy-system/buildings/data-centres-and-data-transmission-networks}
\end{botherref}
\endbibitem

\bibitem[\protect\citeauthoryear{Rappaport et~al.}{2024}]{rappaport24waste}
\begin{barticle}
\bauthor{\bsnm{Rappaport}, \binits{T.S.}},
\bauthor{\bsnm{Ying}, \binits{M.}},
\bauthor{\bsnm{Piovesan}, \binits{N.}},
\bauthor{\bsnm{De~Domenico}, \binits{A.}},
\bauthor{\bsnm{Shakya}, \binits{D.}}:
\batitle{Waste factor and waste figure: A unified theory for modeling and analyzing wasted power in radio access networks for improved sustainability}.
\bjtitle{IEEE Open Journal of the Communications Society}
\bvolume{5},
\bfpage{4839}--\blpage{4867}
(\byear{2024})
\doiurl{10.1109/OJCOMS.2024.3431459}
\end{barticle}
\endbibitem

\bibitem[\protect\citeauthoryear{Rappaport et~al.}{2013}]{rappaport2013millimeter}
\begin{barticle}
\bauthor{\bsnm{Rappaport}, \binits{T.S.}},
\bauthor{\bsnm{Sun}, \binits{S.}},
\bauthor{\bsnm{Mayzus}, \binits{R.}},
\bauthor{\bsnm{Zhao}, \binits{H.}},
\bauthor{\bsnm{Azar}, \binits{Y.}},
\bauthor{\bsnm{Wang}, \binits{K.}},
\bauthor{\bsnm{Wong}, \binits{G.N.}},
\bauthor{\bsnm{Schulz}, \binits{J.K.}},
\bauthor{\bsnm{Samimi}, \binits{M.}},
\bauthor{\bsnm{Gutierrez}, \binits{F.}}:
\batitle{Millimeter wave mobile communications for {5G} cellular: It will work!}
\bjtitle{IEEE access}
\bvolume{1},
\bfpage{335}--\blpage{349}
(\byear{2013})
\end{barticle}
\endbibitem

\bibitem[\protect\citeauthoryear{Shakya et~al.}{2024}]{shakya2024propagation}
\begin{botherref}
\oauthor{\bsnm{Shakya}, \binits{D.}},
\oauthor{\bsnm{Ying}, \binits{M.}},
\oauthor{\bsnm{Rappaport}, \binits{T.S.}},
\oauthor{\bsnm{Poddar}, \binits{H.}},
\oauthor{\bsnm{Ma}, \binits{P.}},
\oauthor{\bsnm{Wang}, \binits{Y.}},
\oauthor{\bsnm{Al-Wazani}, \binits{I.}}:
Propagation measurements and channel models in indoor environment at 6.75 ghz fr1 (c) and 16.95 ghz fr3 upper-mid band spectrum for {5G} and {6G}.
arXiv preprint arXiv:2405.01358
(2024)
\end{botherref}
\endbibitem

\bibitem[\protect\citeauthoryear{Kang et~al.}{2024}]{kang2024cellular}
\begin{botherref}
\oauthor{\bsnm{Kang}, \binits{S.}},
\oauthor{\bsnm{Mezzavilla}, \binits{M.}},
\oauthor{\bsnm{Rangan}, \binits{S.}},
\oauthor{\bsnm{Madanayake}, \binits{A.}},
\oauthor{\bsnm{Venkatakrishnan}, \binits{S.B.}},
\oauthor{\bsnm{Hellbourg}, \binits{G.}},
\oauthor{\bsnm{Ghosh}, \binits{M.}},
\oauthor{\bsnm{Rahmani}, \binits{H.}},
\oauthor{\bsnm{Dhananjay}, \binits{A.}}:
Cellular wireless networks in the upper mid-band.
IEEE Open Journal of the Communications Society
(2024)
\end{botherref}
\endbibitem

\bibitem[\protect\citeauthoryear{Shakya et~al.}{2024}]{shakya2024wideband}
\begin{botherref}
\oauthor{\bsnm{Shakya}, \binits{D.}},
\oauthor{\bsnm{Ying}, \binits{M.}},
\oauthor{\bsnm{Rappaport}, \binits{T.S.}},
\oauthor{\bsnm{Poddar}, \binits{H.}},
\oauthor{\bsnm{Ma}, \binits{P.}},
\oauthor{\bsnm{Wang}, \binits{Y.}},
\oauthor{\bsnm{Al-Wazani}, \binits{I.}}:
Wideband penetration loss through building materials and partitions at 6.75 {GHz} in {FR1} ({C}) and 16.95 {GHz} in the {FR3} upper mid-band spectrum.
arXiv preprint arXiv:2405.01362
(2024)
\end{botherref}
\endbibitem

\bibitem[\protect\citeauthoryear{Polat et~al.}{2025}]{11027214}
\begin{bchapter}
\bauthor{\bsnm{Polat}, \binits{C.}},
\bauthor{\bsnm{Kurban}, \binits{H.}},
\bauthor{\bsnm{Serpedin}, \binits{E.}},
\bauthor{\bsnm{Kurban}, \binits{M.}}:
\bctitle{Data-efficient hydrogen adsorption prediction in copper nanoclusters: A computer vision-based transfer learning approach}.
In: \bbtitle{2025 IEEE 19th International Conference on Compatibility, Power Electronics and Power Engineering (CPE-POWERENG)},
pp. \bfpage{1}--\blpage{6}
(\byear{2025}).
\doiurl{10.1109/CPE-POWERENG63314.2025.11027214}
\end{bchapter}
\endbibitem

\bibitem[\protect\citeauthoryear{Huawei}{2020}]{HuaweiGreen5G}
\begin{botherref}
\oauthor{\bsnm{Huawei}}:
Green {5G}: Building a Sustainable World
(2020).
\url{https://www.huawei.com/en/public-policy/green-5g-building-a-sustainable-world}
\end{botherref}
\endbibitem

\bibitem[\protect\citeauthoryear{Murdock and Rappaport}{2013}]{murdock2013consumption}
\begin{barticle}
\bauthor{\bsnm{Murdock}, \binits{J.N.}},
\bauthor{\bsnm{Rappaport}, \binits{T.S.}}:
\batitle{Consumption factor and power-efficiency factor: A theory for evaluating the energy efficiency of cascaded communication systems}.
\bjtitle{IEEE Journal on Selected Areas in Communications}
\bvolume{32}(\bissue{2}),
\bfpage{221}--\blpage{236}
(\byear{2013})
\end{barticle}
\endbibitem

\bibitem[\protect\citeauthoryear{Kanhere et~al.}{2022}]{kanhere2022power}
\begin{barticle}
\bauthor{\bsnm{Kanhere}, \binits{O.}},
\bauthor{\bsnm{Poddar}, \binits{H.}},
\bauthor{\bsnm{Xing}, \binits{Y.}},
\bauthor{\bsnm{Shakya}, \binits{D.}},
\bauthor{\bsnm{Ju}, \binits{S.}},
\bauthor{\bsnm{Rappaport}, \binits{T.S.}}:
\batitle{A power efficiency metric for comparing energy consumption in future wireless networks in the millimeter-wave and terahertz bands}.
\bjtitle{IEEE Wireless Communications}
\bvolume{29}(\bissue{6}),
\bfpage{56}--\blpage{63}
(\byear{2022})
\end{barticle}
\endbibitem

\bibitem[\protect\citeauthoryear{Chen et~al.}{2010}]{chen2010energy}
\begin{bchapter}
\bauthor{\bsnm{Chen}, \binits{T.}},
\bauthor{\bsnm{Kim}, \binits{H.}},
\bauthor{\bsnm{Yang}, \binits{Y.}}:
\bctitle{Energy efficiency metrics for green wireless communications}.
In: \bbtitle{2010 International Conference on Wireless Communications \& Signal Processing (WCSP)},
pp. \bfpage{1}--\blpage{6}
(\byear{2010}).
\bcomment{IEEE}
\end{bchapter}
\endbibitem

\bibitem[\protect\citeauthoryear{Xing et~al.}{2021}]{xing21high}
\begin{bchapter}
\bauthor{\bsnm{Xing}, \binits{Y.}},
\bauthor{\bsnm{Hsieh}, \binits{F.}},
\bauthor{\bsnm{Ghosh}, \binits{A.}},
\bauthor{\bsnm{Rappaport}, \binits{T.S.}}:
\bctitle{High altitude platform stations ({HAPS}): Architecture and system performance}.
In: \bbtitle{2021 IEEE 93rd Vehicular Technology Conference (VTC2021-Spring)}
(\byear{2021}).
\doiurl{10.1109/VTC2021-Spring51267.2021.9448899}
\end{bchapter}
\endbibitem

\bibitem[\protect\citeauthoryear{von Perner~et al.}{2021}]{NGMN2021}
\begin{botherref}
\oauthor{\bsnm{al.}, \binits{J.}}:
Green Future Networks: Network Energy Efficiency.
NGMN Alliance.
\url{https://www.ngmn.org/wp-content/uploads/211009-GFN-Network-Energy-Efficiency-1.0.pdf}
\end{botherref}
\endbibitem

\bibitem[\protect\citeauthoryear{ETSI Environmental Engineering}{2020a}]{ETSI2020}
\begin{botherref}
{Es2023228}: Assessment of Mobile Network Energy Efficiency.
ETSI Environmental Engineering.
\url{https://www.etsi.org/deliver/etsi_es/203200_203299/203228/01.03.01_60/es_203228v010301p.pdf}
\end{botherref}
\endbibitem

\bibitem[\protect\citeauthoryear{ETSI}{2020b}]{ETSI2020_TS103786}
\begin{botherref}
Environmental Engineering ({EE}): {TS103786}: Measurement Method for Energy Efficiency of Wireless Access Network Equipment; Dynamic Energy Efficiency Measurement Method of {5G} Base Station ({BS}).
ETSI.
\url{{https://www.etsi.org/deliver/etsi_ts/103700_103799/103786/01.01.01_60/ts_103786v010101p.pdf}}
\end{botherref}
\endbibitem

\bibitem[\protect\citeauthoryear{International Telecommunications Union}{2016}]{ITU2016}
\begin{botherref}
Energy Efficiency Measurement for Telecommunication Equipment, {Rec.-ITU-T L.1350-201610}.
International Telecommunications Union.
\url{https://www.itu.int/rec/T-REC-L.1350-201610-I/en}
\end{botherref}
\endbibitem

\bibitem[\protect\citeauthoryear{International Telecommunications Union}{2020}]{ITU2020}
\begin{botherref}
Energy Efficiency Metrics and Measurement Methods for Telecommunication Equipment, {Rec. ITU-L.1310}.
International Telecommunications Union.
\url{https://www.itu.int/rec/T-REC-L.1310-202009-I}
\end{botherref}
\endbibitem

\bibitem[\protect\citeauthoryear{3rd Generation Partnership Project (3GPP)}{2021}]{3GPP2021}
\begin{botherref}
Technical Specification Group Services and System Aspects; Management and Orchestration; Study on New Aspects of Energy Efficiency ({EE}) for {5G}; ({R}elease 17), {V}ersion 17.0.0.
3rd Generation Partnership Project (3GPP).
\url{https://www.3gpp.org}
\end{botherref}
\endbibitem

\bibitem[\protect\citeauthoryear{Zahnstecher et~al.}{2023}]{10520551}
\begin{bchapter}
\bauthor{\bsnm{Zahnstecher}, \binits{B.}},
\bauthor{\bsnm{Carobolante}, \binits{F.}},
\bauthor{\bsnm{McCune}, \binits{E.}},
\bauthor{\bsnm{Gerber}, \binits{D.}},
\bauthor{\bsnm{Kirkpatrick}, \binits{D.}},
\bauthor{\bsnm{Olsson}, \binits{M.}},
\bauthor{\bsnm{Booth}, \binits{R.}},
\bauthor{\bsnm{Björnson}, \binits{E.}},
\bauthor{\bsnm{Bresniker}, \binits{K.}},
\bauthor{\bsnm{Draxler}, \binits{P.}},
\bauthor{\bsnm{Nease}, \binits{L.}},
\bauthor{\bsnm{Darema}, \binits{F.}},
\bauthor{\bsnm{Bandyopadhyay}, \binits{A.}},
\bauthor{\bsnm{Alouini}, \binits{M.-S.}},
\bauthor{\bsnm{Nordman}, \binits{B.}}:
\bctitle{Ingr roadmap energy efficiency chapter}.
In: \bbtitle{2023 IEEE Future Networks World Forum (FNWF)},
pp. \bfpage{1}--\blpage{148}
(\byear{2023}).
\doiurl{10.1109/FNWF58287.2023.10520551}
\end{bchapter}
\endbibitem

\bibitem[\protect\citeauthoryear{Ju et~al.}{2019}]{ju2019scattering}
\begin{bchapter}
\bauthor{\bsnm{Ju}, \binits{S.}},
\bauthor{\bsnm{Shah}, \binits{S.H.A.}},
\bauthor{\bsnm{Javed}, \binits{M.A.}},
\bauthor{\bsnm{Li}, \binits{J.}},
\bauthor{\bsnm{Palteru}, \binits{G.}},
\bauthor{\bsnm{Robin}, \binits{J.}},
\bauthor{\bsnm{Xing}, \binits{Y.}},
\bauthor{\bsnm{Kanhere}, \binits{O.}},
\bauthor{\bsnm{Rappaport}, \binits{T.S.}}:
\bctitle{Scattering mechanisms and modeling for terahertz wireless communications}.
In: \bbtitle{ICC 2019-2019 IEEE International Conference on Communications (ICC)},
pp. \bfpage{1}--\blpage{7}
(\byear{2019}).
\bcomment{IEEE}
\end{bchapter}
\endbibitem

\bibitem[\protect\citeauthoryear{Nie et~al.}{2013}]{6666553}
\begin{bchapter}
\bauthor{\bsnm{Nie}, \binits{S.}},
\bauthor{\bsnm{MacCartney}, \binits{G.R.}},
\bauthor{\bsnm{Sun}, \binits{S.}},
\bauthor{\bsnm{Rappaport}, \binits{T.S.}}:
\bctitle{72 ghz millimeter wave indoor measurements for wireless and backhaul communications}.
In: \bbtitle{2013 IEEE 24th Annual International Symposium on Personal, Indoor, and Mobile Radio Communications (PIMRC)},
pp. \bfpage{2429}--\blpage{2433}
(\byear{2013}).
\doiurl{10.1109/PIMRC.2013.6666553}
\end{bchapter}
\endbibitem

\bibitem[\protect\citeauthoryear{Rappaport et~al.}{2012}]{6175397}
\begin{bchapter}
\bauthor{\bsnm{Rappaport}, \binits{T.S.}},
\bauthor{\bsnm{Qiao}, \binits{Y.}},
\bauthor{\bsnm{Tamir}, \binits{J.I.}},
\bauthor{\bsnm{Murdock}, \binits{J.N.}},
\bauthor{\bsnm{Ben-Dor}, \binits{E.}}:
\bctitle{Cellular broadband millimeter wave propagation and angle of arrival for adaptive beam steering systems (invited paper)}.
In: \bbtitle{2012 IEEE Radio and Wireless Symposium},
pp. \bfpage{151}--\blpage{154}
(\byear{2012}).
\doiurl{10.1109/RWS.2012.6175397}
\end{bchapter}
\endbibitem

\bibitem[\protect\citeauthoryear{Friis}{1944}]{friis1944noise}
\begin{barticle}
\bauthor{\bsnm{Friis}, \binits{H.T.}}:
\batitle{Noise figures of radio receivers}.
\bjtitle{Proceedings of the IRE}
\bvolume{32}(\bissue{7}),
\bfpage{419}--\blpage{422}
(\byear{1944})
\end{barticle}
\endbibitem

\bibitem[\protect\citeauthoryear{Rappaport et~al.}{2024}]{ying2023waste}
\begin{barticle}
\bauthor{\bsnm{Rappaport}, \binits{T.}},
\bauthor{\bsnm{Ying}, \binits{M.}},
\bauthor{\bsnm{Shakya}, \binits{D.}}:
\batitle{Waste figure and waste factor: New metrics for evaluating power efficiency in any circuit or cascade}.
\bjtitle{Microwave Journal}
\bvolume{67}(\bissue{5}),
\bfpage{54}--\blpage{84}
(\byear{2024})
\end{barticle}
\endbibitem

\bibitem[\protect\citeauthoryear{Shannon}{1948}]{shannon1948mathematical}
\begin{barticle}
\bauthor{\bsnm{Shannon}, \binits{C.E.}}:
\batitle{A mathematical theory of communication}.
\bjtitle{The Bell system technical journal}
\bvolume{27}(\bissue{3}),
\bfpage{379}--\blpage{423}
(\byear{1948})
\end{barticle}
\endbibitem

\bibitem[\protect\citeauthoryear{Razavi}{2011}]{razavi2011rf}
\begin{bbook}
\bauthor{\bsnm{Razavi}, \binits{B.}}:
\bbtitle{RF Microelectronics},
\bedition{2nd} edn.
\bpublisher{Pearson}, \blocation{???}
(\byear{2011})
\end{bbook}
\endbibitem

\bibitem[\protect\citeauthoryear{Sweet}{1990}]{sweet1990mic}
\begin{bbook}
\bauthor{\bsnm{Sweet}, \binits{A.A.}}:
\bbtitle{MIC \& MMIC Amplifier and Oscillator Circuit Design}.
\bsertitle{Artech House Microwave Library},
p. \bfpage{380}.
\bpublisher{Artech House Publishers},
\blocation{Norwood, MA}
(\byear{1990})
\end{bbook}
\endbibitem

\bibitem[\protect\citeauthoryear{Dayarathna et~al.}{2015}]{dayarathna2015data}
\begin{barticle}
\bauthor{\bsnm{Dayarathna}, \binits{M.}},
\bauthor{\bsnm{Wen}, \binits{Y.}},
\bauthor{\bsnm{Fan}, \binits{R.}}:
\batitle{Data center energy consumption modeling: A survey}.
\bjtitle{IEEE Communications surveys \& tutorials}
\bvolume{18}(\bissue{1}),
\bfpage{732}--\blpage{794}
(\byear{2015})
\end{barticle}
\endbibitem

\bibitem[\protect\citeauthoryear{Friis}{1944}]{noise_figure}
\begin{barticle}
\bauthor{\bsnm{Friis}, \binits{H.T.}}:
\batitle{Noise figures of radio receivers}.
\bjtitle{Proceedings of the IRE}
\bvolume{32}(\bissue{7}),
\bfpage{419}--\blpage{422}
(\byear{1944})
\doiurl{10.1109/JRPROC.1944.232049}
\end{barticle}
\endbibitem

\bibitem[\protect\citeauthoryear{Sklar}{2021}]{sklar2021digital}
\begin{bbook}
\bauthor{\bsnm{Sklar}, \binits{B.}}:
\bbtitle{Digital Communications: Fundamentals and Applications}.
\bpublisher{Pearson}, \blocation{???}
(\byear{2021})
\end{bbook}
\endbibitem

\bibitem[\protect\citeauthoryear{Cover}{1999}]{cover1999elements}
\begin{bbook}
\bauthor{\bsnm{Cover}, \binits{T.M.}}:
\bbtitle{Elements of Information Theory}.
\bpublisher{John Wiley \& Sons}, \blocation{???}
(\byear{1999})
\end{bbook}
\endbibitem

\bibitem[\protect\citeauthoryear{}{2024}]{ericsson2024mobility}
\begin{botherref}
{Ericsson Mobility Report: November 2024}.
Accessed: 2025-03-30.
\url{https://www.ericsson.com/4adb7e/assets/local/reports-papers/mobility-report/documents/2024/ericsson-mobility-report-november-2024.pdf}
\end{botherref}
\endbibitem

\bibitem[\protect\citeauthoryear{Ying et~al.}{2024}]{ying_waste_globecom_24}
\begin{bchapter}
\bauthor{\bsnm{Ying}, \binits{M.}},
\bauthor{\bsnm{Shakya}, \binits{D.}},
\bauthor{\bsnm{Rappaport}, \binits{T.S.}}:
\bctitle{Using waste factor to optimize energy efficiency in multiple-input single-output (miso) and multiple-input multiple-output (mimo) systems}.
In: \bbtitle{GLOBECOM 2024 - 2024 IEEE Global Communications Conference},
pp. \bfpage{3170}--\blpage{3175}
(\byear{2024}).
\doiurl{10.1109/GLOBECOM52923.2024.10900955}
\end{bchapter}
\endbibitem

\end{thebibliography}

\includepdf[pages=-]{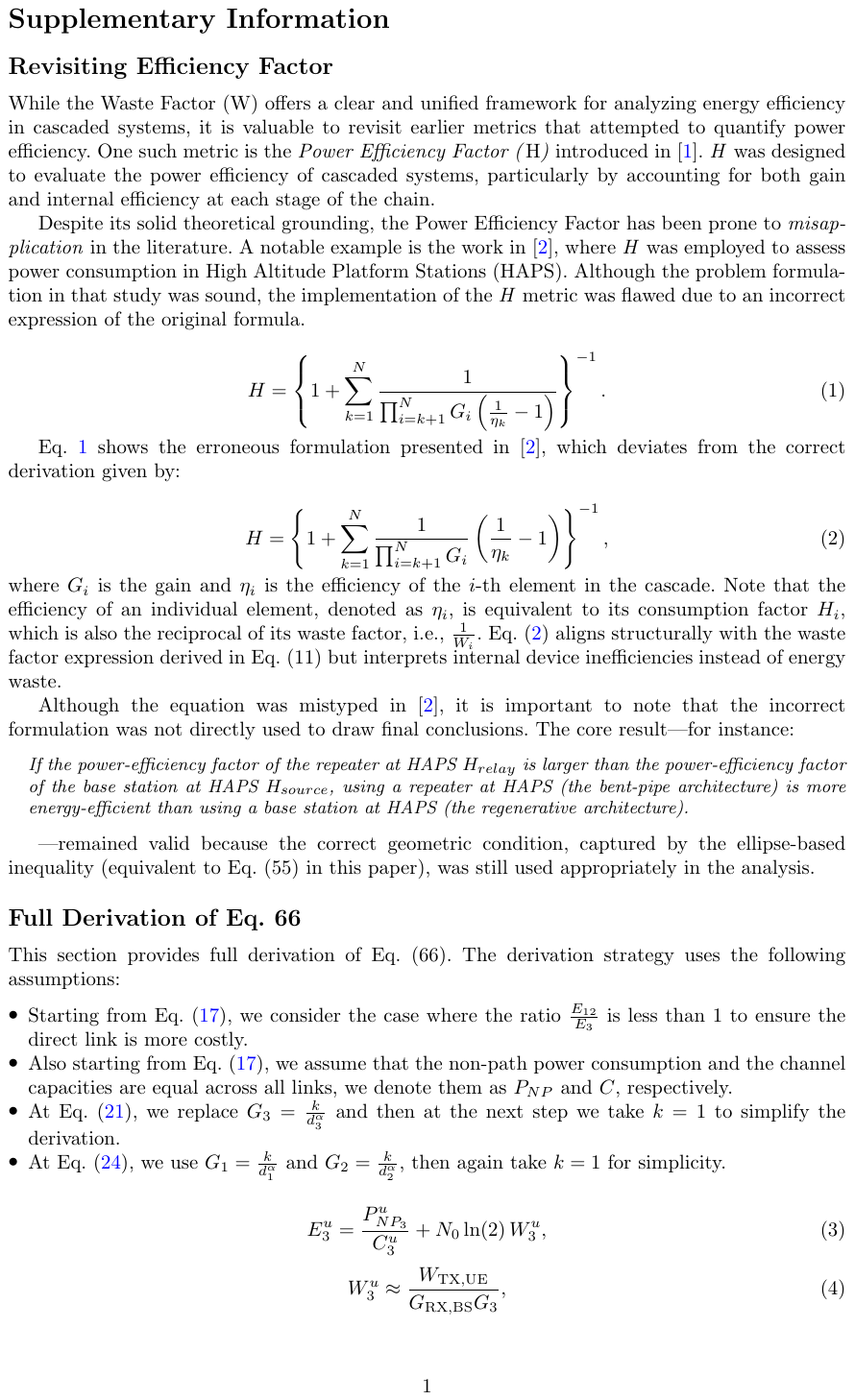}
\end{document}